\documentclass[twocolumn]{aastex61}

\usepackage{amssymb}
\usepackage{amsmath}
\usepackage{latexsym}
\usepackage{fnpos}
\usepackage{scrextend}
\usepackage{scalerel}
\usepackage{hyperref}
\usepackage{tabularx}
\usepackage{xspace}
\usepackage{enumerate}
\usepackage{graphicx}
\usepackage{url}
\usepackage{longtable}
\usepackage{rotating}
\newcommand{\eal}[2]{\ifmmode{\mathrm{#1\,#2}}\else{#1\textsc{$\,$\lowercase{#2}}}\fi\xspace}
\newcommand{\feal}[2]{\ifmmode{\mathrm{#1\,#2}}\else{[#1\textsc{$\,$\lowercase{#2}}]}\fi\xspace}
\newcommand{\hfeal}[2]{\ifmmode{\mathrm{#1\,#2}}\else{#1\textsc{$\,$\lowercase{#2}}]}\fi\xspace}

\received{***}
\revised{***}
\accepted{***}
\submitjournal{ApJ}

\shorttitle{Nova ASASSN-17\MakeLowercase{pf}}
\shortauthors{Aydi et al.}

\begin{document}

\title{Flaring, dust formation, and shocks in the very slow nova ASASSN-17\MakeLowercase{pf} (LMCN 2017-11\MakeLowercase{a})}

\correspondingauthor{Elias Aydi}
\email{aydielia@pa.msu.edu}

\author[0000-0001-8525-3442]{E.~Aydi}
\affil{Center for Data Intensive and Time Domain Astronomy, Department of Physics and Astronomy, Michigan State University, East Lansing, MI 48824, USA \\}

\author{L.~Chomiuk}
\affiliation{Center for Data Intensive and Time Domain Astronomy, Department of Physics and Astronomy, Michigan State University, East Lansing, MI 48824, USA \\}

\author{J.~Strader}
\affiliation{Center for Data Intensive and Time Domain Astronomy, Department of Physics and Astronomy, Michigan State University, East Lansing, MI 48824, USA \\}

\author{S.~J.~Swihart}
\affiliation{Center for Data Intensive and Time Domain Astronomy, Department of Physics and Astronomy, Michigan State University, East Lansing, MI 48824, USA \\}

\author{A.~Bahramian}
\affiliation{International Centre for Radio Astronomy Research, Curtin University, GPO Box U1987, Perth, WA 6845, Australia \\}
\affiliation{Center for Data Intensive and Time Domain Astronomy, Department of Physics and Astronomy, Michigan State University, East Lansing, MI 48824, USA \\}

\author[0000-0002-3014-3665]{E.~J.~Harvey} 
\affiliation{Astrophysics Research Institute, Liverpool John Moores University, Liverpool, L3 5RF, UK \\}

\author{C.~T.~Britt}
\affiliation{Space Telescope Science Institute, 3700 San Martin Drive, Baltimore, MD, 21218, USA \\}

\author{D.~A.~H.~Buckley}
\affiliation{South African Astronomical Observatory, P.O. Box 9, 7935 Observatory, South Africa \\}

\author{Ping Chen}
\affiliation{Kavli Institute for Astronomy and Astrophysics, Peking University, Yi He Yuan Road 5, Hai Dian District, Beijing 100871, China \\}

\author{K.~Dage}
\affiliation{Center for Data Intensive and Time Domain Astronomy, Department of Physics and Astronomy, Michigan State University, East Lansing, MI 48824, USA \\}

\author[0000-0003-0156-3377]{M.~J.~Darnley}
\affiliation{Astrophysics Research Institute, Liverpool John Moores University, Liverpool, L3 5RF, UK \\}

\author{S.~Dong}
\affiliation{Kavli Institute for Astronomy and Astrophysics, Peking University, Yi He Yuan Road 5, Hai Dian District, Beijing 100871, China \\}

\author{F-J.~Hambsch}
\affiliation{Vereniging Voor Sterrenkunde (VVS), Oude Bleken 12, 2400 Mol, Belgium \\}
\affiliation{Bundesdeutsche Arbeitsgemeinschaft f$\ddot{u}$r Ver$\ddot{a}$nderliche Sterne (BAV), Munsterdamm 90, 12169 Berlin, Germany \\}

\author[0000-0001-9206-3460]{T.~W.-S.~Holoien}
\affiliation{The Observatories of the Carnegie Institution for Science, 813 Santa Barbara St., Pasadena, CA 91101, USA\\}

\author[0000-0001-8738-6011]{S.~W.~Jha}
\affiliation{Department of Physics and Astronomy, Rutgers, the State University of New Jersey, 136 Frelinghuysen Road, Piscataway, NJ 08854, USA \\}

\author{C.~S.~Kochanek}
\affiliation{Department of Astronomy, The Ohio State University, 140 West 18th Avenue, Columbus, OH 43210, USA \\}

\author{N.~P.~M.~Kuin}
\affiliation{Mullard Space Science Laboratory, Dept. of Space and Climate Sciences, University College London, Holmbury St Mary, Dorking, RH5 6NT, UK \\}

\author{K.~L.~Li}
\affiliation{Department of Physics, UNIST, Ulsan 44919, Korea \\}

\author{L.~A.~G.~Monard}
\affiliation{Kleinkaroo Observatory, Calitzdorp, South Africa \\}

\author{K.~Mukai}
\affiliation{CRESST and X-ray Astrophysics Laboratory, NASA/GSFC, Greenbelt, MD 20771, USA \\}
\affiliation{Department of Physics, University of Maryland, Baltimore County, 1000 Hilltop Circle, Baltimore, MD 21250, USA \\}

\author[0000-0001-5624-2613]{K.~L. Page}
\affiliation{X-ray and Observational Astronomy Group, Department of Physics and Astronomy, University of Leicester, LE1 7RH, UK \\}

\author{J.~L.~Prieto}
\affiliation{N\'ucleo de Astronom\'ia de la Facultad de Ingenier\'ia y Ciencias, Universidad Diego Portales, Av. Ej\'ercito 441, Santiago, Chile \\}
\affiliation{Millennium Institute of Astrophysics, Santiago, Chile \\}

\author[0000-0002-2806-9339]{N.~D.~Richardson}
\affiliation{Ritter Observatory, Department of Physics and Astronomy, The University of Toledo, Toledo, OH 43606-3390, USA \\}

\author[0000-0003-4631-1149]{B.~J.~Shappee}
\affiliation{Institute for Astronomy, University of Hawai'i, 2680 Woodlawn Drive, Honolulu, HI 96822, USA \\}

\author[0000-0003-0286-7858]{L.~Shishkovsky}
\affiliation{Center for Data Intensive and Time Domain Astronomy, Department of Physics and Astronomy, Michigan State University, East Lansing, MI 48824, USA \\}

\author[0000-0001-5991-6863]{K.~V.~Sokolovsky}
\affiliation{Center for Data Intensive and Time Domain Astronomy, Department of Physics and Astronomy, Michigan State University, East Lansing, MI 48824, USA \\}
\affiliation{Sternberg Astronomical Institute, Moscow State University, Universitetskii~pr.~13, 119992~Moscow, Russia \\}
\affiliation{Astro Space Center of Lebedev Physical Institute, Profsoyuznaya~St.~84/32, 117997~Moscow, Russia \\}

\author{K.~Z.~Stanek}
\affiliation{Department of Astronomy, The Ohio State University, 140 West 18th Avenue, Columbus, OH 43210, USA \\}

\author{T.~Thompson}
\affiliation{Department of Astronomy, The Ohio State University, 140 West 18th Avenue, Columbus, OH 43210, USA \\}



\begin{abstract}

We present a detailed study of the 2017 eruption of the classical nova ASASSN-17pf (LMCN 2017-11a), which is located in the Large Magellanic Cloud, including data from AAVSO, ASAS-SN, SALT, SMARTS, SOAR, and the Neil Gehrels \textit{Swift} Observatory. The optical light-curve is characterized by multiple maxima (flares) on top of a slowly evolving light-curve (with a decline time, $t_2>$ 100\,d). 
The maxima correlate with the appearance of new absorption line systems in the optical spectra characterized by increasing radial velocities. We suggest that this is evidence of multiple episodes of mass-ejection with increasing expansion velocities. The line profiles in the optical spectra indicate very low expansion velocities (FWHM $\sim$ 190\,km\,s$^{-1}$), making this nova one of the slowest expanding ever observed, consistent with the slowly evolving light-curve. The evolution of the colors and spectral energy distribution show evidence of decreasing temperatures and increasing effective radii for the pseudo-photosphere during each maximum. The optical and infrared light-curves are consistent with dust formation 125 days post-discovery. We speculate that novae showing several optical maxima have multiple mass-ejection episodes leading to shocks that may drive $\gamma$-ray emission and dust formation.
\end{abstract}

\keywords{stars: individual (ASASSN-17pf) --- novae, cataclysmic variables --- white dwarfs.}



\section{Introduction}

Classical novae (CNe) are transient events driven by runaway thermonuclear reactions
on the surfaces of accreting white dwarfs (WDs) in interacting binary systems (see \citealt{Bode_etal_2008} for a general review; also \citealt{Starrfield_etal_2008,2016PASP..128e1001S}). This leads to an increase in brightness by 8\,--\,15\,mag and ejection of material with velocities ranging between a few hundred and a many thousand km\,s$^{-1}$ \citep{Payne-Gaposchkin_1957,Gallaher_etal_1978}.

The nova rate in the Large Magellanic Cloud (LMC) is estimated to be $\sim$ 2 eruptions per year \citep{Mroz_etal_2016}. While LMC novae are relatively rare, they are of  great interest due to the known distance ($d\approx 50.0 \pm 2.0$\;kpc; \citealt{Pietrzyski_etal_2013}) and the low Galactic extinction towards the LMC. This allows one to accurately determine physical properties of the eruption compared to Galactic novae for which accurate determination of distance and reddening is challenging.

Nova ASASSN-17pf (LMCN 2017-11a) was discovered by the All-Sky Automated Survey for Supernovae (ASAS-SN; \citealt{Shappee_etal_2014,Kochanek_etal_2017}) on HJD 2458074.7 (2017 November 17.2 UT) at $V$ = 13.8. The nova is located in the LMC with equatorial coordinates of $(\alpha, \delta)_{\rm J2000.0}$ = (5$^{\mathrm{h}}$30$^{\mathrm{m}}$16$^{\mathrm{s}}$\!.80, --73$^{\circ}$16$'$11\arcsec\!\!.0). In the following we will assume $t_0$ = HJD 2458074.7 as the start of the eruption. The last pre-eruption observation obtained of the system by ASAS-SN was on HJD 2458058.7 and therefore, the eruption may have started earlier than HJD 2458074.7.

 Many novae deviate from the stereotypical smoothly declining optical light-curve (see, e.g., \citealt{Strope_etal_2010}). One intriguing behavior is the appearance of multiple flares or maxima on top of a slowly evolving light-curve. Several explanations have been suggested to explain these flares, such as pulsations/instabilities in the envelope of the WD \citep{Schenker_1999,Pejcha_2009} possibly leading to multiple ejection episodes \citep{Cassatella_etal_2004,Csak_etal_2005,Hillman_etal_2014}, instabilities in a massive accretion disk that survived the eruption \citep{Goranskij_etal_2007}, or mass transfer bursts from the secondary to the WD \citep{Chocol_Pribulla_1998}. No preferred model to explain the optical flares has emerged, either observationally or theoretically. A central goal of this paper is to better understand the conditions in novae during optical flares.

In this paper we present optical, near-infrared (NIR), and near-ultraviolet (NUV) observations of ASASSN-17pf, which is one of the slowest expanding novae known and also shows repeated flares in its optical light-curve. The paper is outlined as follows:\ in Section~\ref{sec_obs} we present the observations. The results are presented in Section~\ref{sec_results} followed by our discussion in Section~\ref{sec_disc}. We present a summary and the conclusions in Section~\ref{sec_conc}.  

\section{Observations}
\label{sec_obs}
\subsection{Optical and near-infrared photometry}
We monitored the eruption of ASASSN-17pf using ANDICAM mounted on the 1.3-m telescope of the Small and Moderate Aperture Research Telescope System (SMARTS) at the Cerro Tololo Inter-American Observatory (CTIO) in Chile to obtain optical \textit{BVRI} and near-IR \textit{JHK} photometric measurements over 52 nights between 2017 Dec 06 (day 19) and 2018 May 24 (day 178). The data were reduced following the procedures detailed in \citet{Walter_etal_2012} and \citet{Swihart_etal_2018}. Our final dataset includes 52 photometric measurements in \textit{VRIJH}, 48 in $B$, and 45 in $K$. We provide the full sample of our SMARTS photometry in Table~\ref{table:SMARTS}.

We also obtained broadband photometry at the Kleinkaroo Observatory, Calitzdorp, South Africa using a 35\,cm f/8 Meade RC400 telescope. The images were dark subtracted, flat fielded, and stacked using standard methods. The magnitudes (see Table~\ref{table:berto} for a log of the observations) were derived by differential aperture photometry to UCAC4 \citep{Zacharias_etal_2012,Zacharias_etal_2013} calibrated reference stars.  

We also use optical photometry from ASAS-SN\footnote{\url{https://asas-sn.osu.edu/}} and the American Association of Variable Star
Observers (AAVSO)\footnote{\url{https://www.aavso.org/}}. The ASAS-SN photometry was performed as described in \citet{Kochanek_etal_2017}. These data provide \textit{BVRI}, Visual (Vis.), $g$ band, and clear reduced to $V$ sequence ($CV$) monitoring starting from 2017 November 19 ($\sim$ day 2) and covering the different stages of the eruption.

\subsection{\textit{Swift} observations}
\label{sec_swift}
We monitored ASASSN-17pf with the Neil Gehrels \textit{Swift} Observatory (hereafter, \textit{Swift}; \citealt{Gehrels_etal_2004}) quasi-weekly since shortly after its discovery in 2017 November until 2018 November (the observations are tabulated in Table~\ref{table:swift}). These observations were performed with the X-ray Telescope (XRT; \citealt{Burrows_etal_2005}) in photon counting mode, and in most epochs the Ultra-violet/Optical Telescope (UVOT; \citealt{Roming_etal_2005}) observed the target either with equal exposures in all 6 filters, or in the three UV filters\footnote{see \url{http://www.swift.ac.uk/analysis/uvot/filters.php}} (see Table~\ref{table:UVOT} for a log of the UVOT photometry). Three sets of exposures with the UVOT UV grism were obtained in January, February, and March 2018. They were processed and calibrated as described in \cite{Kuin_etal_2015}, but unfortunately the February spectrum could not be used due to contamination from a bright nearby star.

We used \textit{uvotproduct} to process all the UVOT photometric observations. The photometry was done using a 5$''$ radius for the source aperture and we obtained the background from an annulus region while masking nearby sources inside the annulus. The UVOT photometry calibration has been described in \cite{Poole_etal_2008} and \cite{Breeveld_et_al_2011}. 

We also reduced and analyzed all the XRT data (following standard \textit{Swift}/XRT procedures) to search for an X-ray counterpart. We did not detect an X-ray counterpart for ASASSN-17pf in \textit{Swift}/XRT, with an upper limit of $2.2\times10^{-4}$ count s$^{-1}$ (3-sigma upper limit, over 0.3\,--\,10\,keV) based on all the 65.8 ks of data collected.

\subsection{Optical Spectroscopy}
We obtained optical spectra of ASASSN-17pf with a diversity of telescopes spanning 10--322 days after discovery, as recorded in Table~\ref{table:spec_log}.

\subsubsection{SOAR low- and medium-resolution spectroscopy}

We performed optical spectroscopy of the nova between 2017 November 27 and 2018 August 14 (days 10 to 270; Table~\ref{table:spec_log}) using the Goodman spectrograph \citep{Clemens_etal_2004} on the 4.1\,m Southern Astrophysical Research (SOAR) telescope in Chile. We obtained a single exposure of 300\,s, using a 400 l\,mm$^{-1}$ grating and a 0\arcsec\!\!.95 slit to provide a resolution $R \sim$ 1000 over the range of 3800\,--\,7800\,$\mathrm{\AA}$. We also obtained medium-resolution spectra over days 86 to 270, each of 1200\,s exposure. For these observations we used a 2100 l\,mm$^{-1}$ grating and a 0\arcsec\!\!.95 slit, yielding a resolution $R \sim 5000$ over a range of 4500\,--\,5100\,$\mathrm{\AA}$.

The spectra were reduced using the Image Reduction and Analysis Facility (IRAF; \citealt{Tody_1986}). The wavelength calibration was performed using an Fe lamp and a relative flux calibration has been applied. In Fig.~\ref{Fig:SOAR_400c2} we present the SOAR low-resolution spectra, and in Fig.~\ref{Fig:SOAR_2100} we present the medium-resolution spectra.

\subsubsection{Magellan MIKE and MagE echelle spectroscopy}

We used the Magellan Inamori Kyocera Echelle (MIKE) spectrograph mounted on the Magellan Clay telescope \citep{Bernstein_etal_2003,Shectman_Johns_2003} at the Las Campanas Observatory (LCO) in Chile to obtain an echelle spectrum on the night of 2017 December 03 (day 16). A 2400\,s exposure spectrum was obtained using the MIKE-red instrument and 1\arcsec\!\!.0 slit to provide a resolution $R \sim$ 22,000 over the range of 4850\,--\,9400\,$\mathrm{\AA}$. The spectrum (Fig.~\ref{Fig:mage_spec}) was reduced using the MIKE Pipeline \citep{Kelson_etal_2000,Kelson_2003}.

We also used the \textit{Magellan} Echelle (MagE) spectrograph \citep{Marshall_etal_2008} mounted on the \textit{Magellan} Baade telescope to obtain two echelle spectra on the nights of 2017 December 15 and 30 (days 28 and 43). The spectra (Fig.~\ref{Fig:mage_spec}) were of 600\,s exposures using the 1\arcsec\!\!.0 slit to provide a resolution $R \sim$ 4100 over the range of 3500\,--\,9500\,$\mathrm{\AA}$. The spectra were reduced using the MagE Pipeline \citep{Kelson_etal_2000,Kelson_2003}. 

\subsubsection{Ir$\acute{\mathrm{e}}$n$\acute{\mathrm{e}}$e Du Pont echelle spectroscopy}

We performed high-resolution optical spectroscopy using the Echelle Spectrograph mounted on the 2.5\,m Ir$\acute{\mathrm{e}}$n$\acute{\mathrm{e}}$e Du Pont Telescope\footnote{\url{http://www.lco.cl/telescopes-information/lco/telescopes-information/irenee-du-pont/instruments}} at LCO in Chile. The observations were carried out on the nights of 2017 December 10 and 11 (days 23 and 24) consisting of a 1200\,s exposures each. The 1$''$\!\!\!.\,5 
slit was used to provide a spectral range of 3600\,--\,9500\,$\mathrm{\AA}$ at a resolution $R \sim$ 27,000. The spectra were reduced using standard methods. 

We also used the Wide Field Reimaging CCD Camera (WFCCD) instrument mounted on the Ir$\acute{\mathrm{e}}$n$\acute{\mathrm{e}}$e Du Pont Telescope to obtain a low-resolution 600\,s exposure spectrum on 2018 January 13 (day 57). WFCCD was used in multi-grism mode to provide a range of 3800\,--\,9500\,$\mathrm{\AA}$ at a resolution R $\lesssim$ 1000 (Fig.~\ref{Fig:chiron_spec}).

\subsubsection{CHIRON echelle spectroscopy}

We obtained high-resolution optical spectroscopy using the CHIRON echelle spectrograph \citep{Tokovinin_etal_2013} mounted on the 1.5\,m telescope at the CTIO in Chile, which is operated by the SMARTS \footnote{The archive of optical spectra is available at \url{http://www.astro.sunysb.edu/fwalter/SMARTS/NovaAtlas/}} Consortium on the nights of 2017 December 19 and 28 (days 32 and 41). CHIRON was used in the fibre mode to provide a resolution $R \sim$ 25,000 over the range of 4100\,--\,8900\,$\mathrm{\AA}$. Both spectra (Fig.~\ref{Fig:chiron_spec}) were of 1800\,s exposure and were reduced using IRAF.

\subsubsection{SALT spectropolarimetry}

We obtained spectropolarimetric observations of the nova on the nights of 2017 December 22 and 2018 October 05 (days 35 and 322) using the Robert Stobie Spectrograph (RSS; \citealt{Burgh_etal_2003}; \citealt{Kobulnicky_etal_2003}), mounted on the  Southern African Large Telescope (SALT; \citealt{Buckley_etal_2006}; \citealt{Odonoghue_etal_2006}) situated at the SAAO, Sutherland, South Africa. The RSS is capable of imaging polarimetry and spectropolarimetry (for more details on the optics, see \citealt{Nordsieck_etal_2003}). 
For the first observations, RSS long slit mode was used with a 1\arcsec\!\!.5 slit and the PG900 grating, resulting in a resolution R $\sim$ 900 over the spectral range  4200\,--\,7200\,$\mathrm{\AA}$. For the second observation, the PG300 grating was used, resulting in a resolution of R $\sim$ 400 over a spectral range 3300\,--\,9800\,$\mathrm{\AA}$. Calibration was performed using an Ar arc spectrum taken immediately after the science frames. The spectropolarimetric data reduction was carried out using the PolSALT pipeline (Crawford et al. in preparation). 

\section{Results and analysis}
\label{sec_results}
\subsection{UV, Optical, and IR light-curves}
\label{optical_LC_sec}

\begin{figure*}
\begin{center}
  \includegraphics[width=\textwidth]{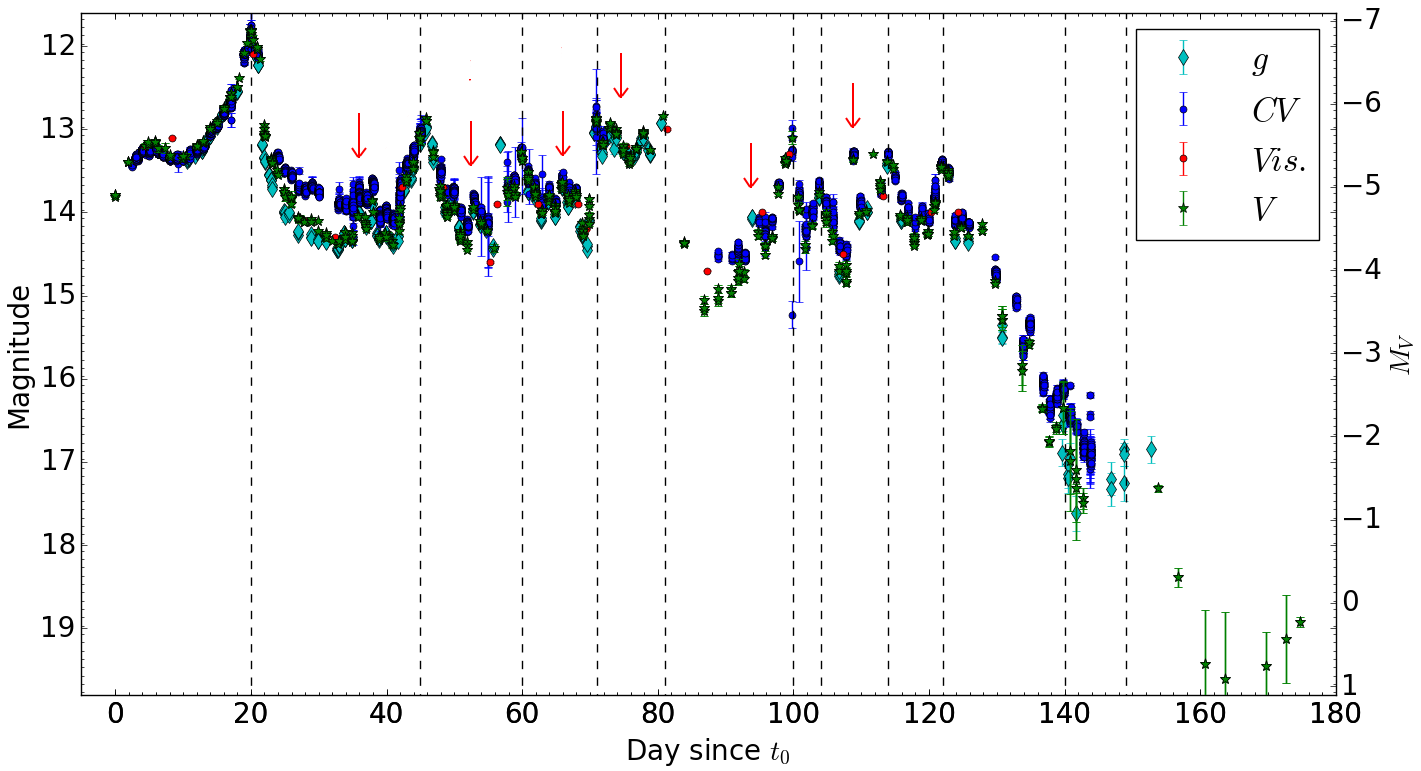}
  \includegraphics[width=\textwidth]{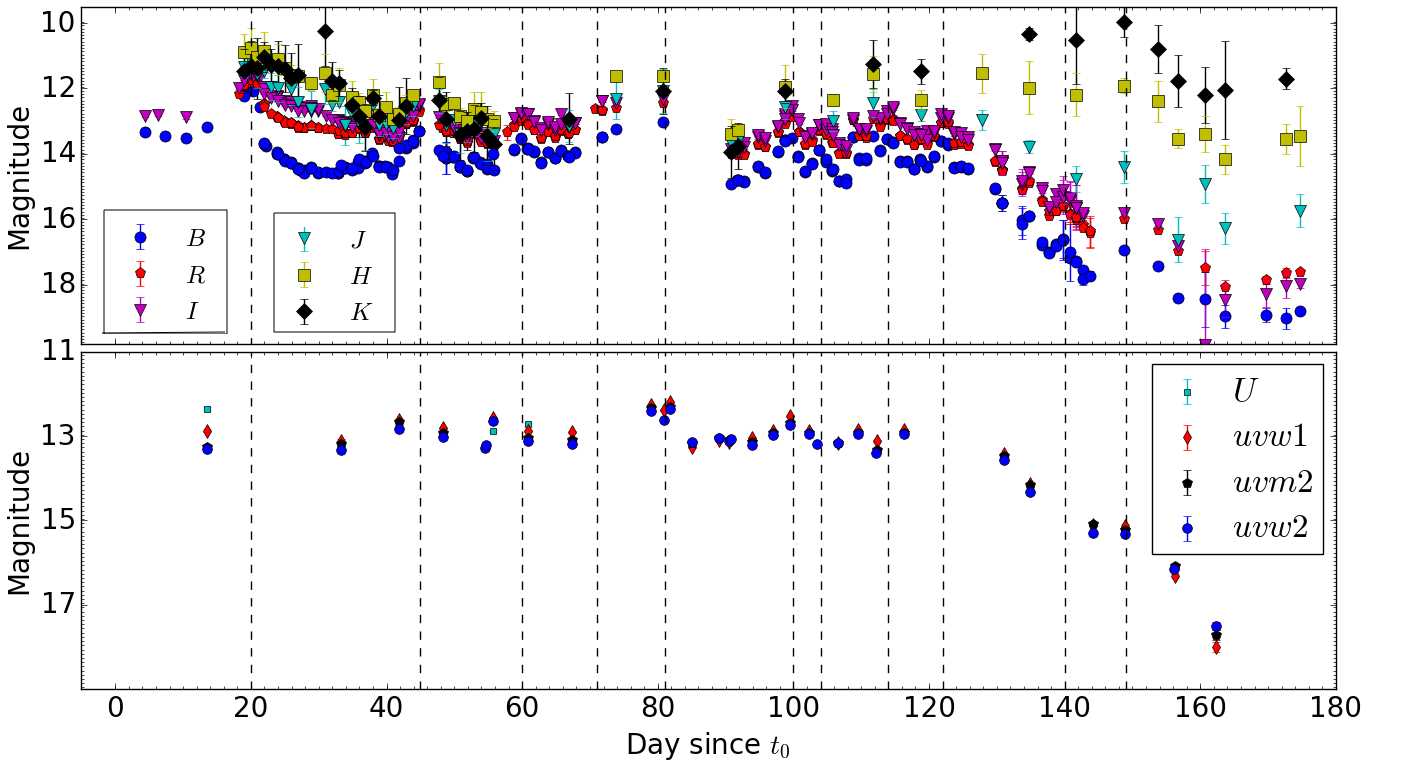}
\caption{The high-cadence visual (top), multi-band optical/near-IR (middle), and NUV (bottom) light-curves of ASASSN-17pf 
plotted relative to the discovery epoch $t_0$. The black vertical dashed lines represent the dates of the maxima. The red arrows mark the position of mini-flares in the light-curve occurring between the maxima (see text for more details). Note that the difference between the ($V$; $g$) bands compared to the ($CV$; visual) measurements during minima is possibly due to the contribution of the emission lines (particularly H$\alpha$) to the unfiltered measurements.}
\label{Fig:VCVVis_LC}
\end{center}
\end{figure*}

Three days after discovery, the magnitude of the nova reached $\approx$ 13.4\,mag in the $V$ band (Fig.~\ref{Fig:VCVVis_LC}). The light-curve then showed a pre-maximum halt (see e.g., \citealt{Hachisu_Kato_2004,Hounsell_etal_2010}) until day $\sim$ 10 after the discovery. A pre-maximum halt typically occurs two magnitudes below maximum \citep{Warner_2008} and may last from a few hours for fast evolving novae up to several months for very slow novae (see e.g., \citealt{Tanaka_etal_2011_159,Tanaka_etal_2011}). Next, the brightness of the system slowly increased for 10 days to reach a maximum of $V_{\mathrm{max}} \approx$ 11.8 mag on HJD 2458094.7 (corresponding to day 20). We consider this date as the time of maximum light ($t_{\mathrm{max}}$) in the $V$ band. 

After the first maximum, the light-curve shows at least 10 additional maxima of different amplitudes and timescales, which are marked in Fig.~\ref{Fig:VCVVis_LC} by the black vertical dashed lines. These are similar to the jitters described in \citet{Strope_etal_2010}, which have been seen in a few other novae such as HR Del \citep{Terzan_1970}, V723 Cas \citep{Munari_etal_1996,Goranskij_etal_2007}, V1548 Aql \citep{Kato_Takamizawa_2001}, V1280 Sco \citep{Hounsell_etal_2010}, and V5558 Sgr \citep{Tanaka_etal_2011}. 
The amplitudes, timescales, and energetics of the optical maxima in ASASSN-17pf are cataloged in  Table~\ref{table:flares}. The main flaring phase lasts until day 125.
It is worth noting that between the maxima, the light-curve exhibits mini-flares with amplitudes $\lesssim$ 0.5\,mag and lasting for a few days, which are marked by red arrows on Fig.~\ref{Fig:VCVVis_LC}. 

\begin{deluxetable}{lcccccc}
\tabletypesize{\small}
\tablewidth{0 pt}
\tablecaption{Characteristics of Optical Maxima
\label{table:flares}}
\tablehead{Nr. & $t_{\mathrm{peak}}$ & $\Delta t$\tablenotemark{a} & $\Delta{V}$\tablenotemark{b} & $\dot{m}_{V,\mathrm{rise}}$ & $\dot{m}_{V,\mathrm{decline}}$ & $\log(E)$\tablenotemark{c}\\ 
 & (d) & (d) & (mag) & (mag\,d$^{-1}$) & (mag\,d$^{-1}$) & (erg)}
\startdata
1 & 20 & 13 & 1.6 & 0.16 & 0.53 & 221.1\\
2 & 45 & 10 & 1.6 & 0.53 & 0.23 & 168.1\\
3 & 60 & 8 & 1.25 & 0.31 & 0.31 & 133.6\\
4 & 71 & --\tablenotemark{d} & 1.5 & 1.5 & --  &  --\\
5 & 81 & --\tablenotemark{d} & 1.5 & -- & 0.5  & --\\
6 & 100 & 15 & 2.1 & 0.16 & 0.65 & 251.6\\
7 & 104 & 6 & 1.0 & 0.5 & 0.25 & 99.8\\
8 & 114 & 9 & 1.4 & 0.23 & 0.35 & 150.4\\
9 & 122 & 7 & 1.0 & 0.25 & 0.25 & 116.4\\
10 & 140 & -- & -- & -- & -- & -- \\
11 & 149 & -- & -- & -- & -- & -- \\
\enddata
\tablenotetext{a}{$\Delta t$ is the duration between the start of the rise and the time when the brightness drops to the same level before the rise (or the time when a new rise starts).}
\tablenotetext{b}{$\Delta(V)$ is the difference between the magnitude at the rise and the magnitude at the peak.}
\tablenotetext{c}{$E$ is the estimated energy emitted during a flare and is equal to $0.5 \times \Delta t \times \Delta (\lambda \L_{\lambda})$; where $\Delta (\lambda \L_{\lambda})$ is the difference between the luminosity at the peak of the flare and the luminosity at the base of the flare.}
\tablenotetext{d}{The sudden rise and drop of the brightness during flares 4 and 5 and the presence of several mini-flares between these two maxima make it difficult to estimate some of their characteristics. These two maxima might also be one long lasting maximum.}
\end{deluxetable}

After the ninth maximum (day $\sim$ 122), ASASSN-17pf showed a fast decline in optical brightness combined with an increase in the NIR  flux (see Fig.~\ref{Fig:VCVVis_LC}). This can be attributed to the obscuration of the central source by the condensation of a dust cloud, where the optical light is absorbed by the dust grains and re-emitted in the IR. Two possible flares can be seen during the dust dip around days 140 and 150 (Fig.~\ref{Fig:VCVVis_LC}).
The extinction event continued until day 164 when the optical brightness reached a minimum ($V \sim$ 19.5); subsequently, it rebrightened by more than 2\,mag over the next $\sim$ hundred days (Fig:~\ref{Fig:base_fit}).

\citet{Pejcha_2009} and \citet{Tanaka_etal_2011_159} have studied several slow novae showing multiple maxima and they concluded that the time interval between maxima follows an increasing systematic trend. In particular, \citet{Pejcha_2009} found that the spacing between the flares is constant on a logarithmic scale and that the time between two consecutive flares follows a power-law of the form:
$$\log (t_i - t_{i-1}) = a + b \log (t_i - t_{\mathrm{max}})\,$$
with $b = 1$, for most of their novae sample. One can see from Fig.~\ref{Fig:time_in_Lc} that the peak spacing of ASASSN-17pf is essentially random (similar to the J-class novae of \citealt{Strope_etal_2010}) and the time intervals do not follow the same power-law of Pejcha's nova sample. Similarly the duration of the flares and the energy emitted during each flare (Table~\ref{table:flares}) show no particular trend, e.g., neither decreasing nor increasing with time.


The speed class of novae is usually described by $t_2$, the time to decline by 2\,mag from maximum light \citep{Payne-Gaposchkin_1957}.
Due to the presence of the flares in the light-curve we use different approaches to estimate $t_2$. One approach is to consider the maximum light as the brightest point of the light-curve ($V_{\mathrm{max}} =$ 11.8) and $t_2$ as the first/last time in the light-curve for which the brightness is below 2\, mag from the peak (see, e.g., \citealt{Strope_etal_2010}). This results in $t_2 \approx 6 \pm 1$ d and $t_2 \approx 103 \pm 1$ d, respectively. Another approach is to ignore the flares and only use the underlying light-curve \citep{Burlak_Henden_2008,Strope_etal_2010}. In this case we consider the maximum magnitude of the light-curve base at $V_{\mathrm{peak, base}}$ = 13.2 on day $t_{\mathrm{peak, base}}$ = 5 and we fit a power law to the data points that constitute only the base (see Fig~\ref{Fig:base_fit}), leading to $t_2 \approx 121 \pm 5$\,d. 

The smallest value of $t_2 \approx$ 6\,d indicates that the nova is very fast---however the larger values both indicate a slow evolving light-curve based on the classification of \citep{Payne-Gaposchkin_1957}, with a mean decline rate $\dot{m}_V \approx$ 0.017 $\pm$ 0.003 mag\,d$^{-1}$. The longer values of $t_2$ seem more representative of the evolution of the light-curves. Such slowly evolving light-curves are often interpreted as a nova taking place on the surface of a low-mass WD ($M_{\mathrm{WD}} \lesssim $1.0\,M$_{\odot}$; see, e.g., \citealt{Yaron_etal_2005}).  

\begin{figure*}
\begin{center}
  \includegraphics[width=\textwidth]{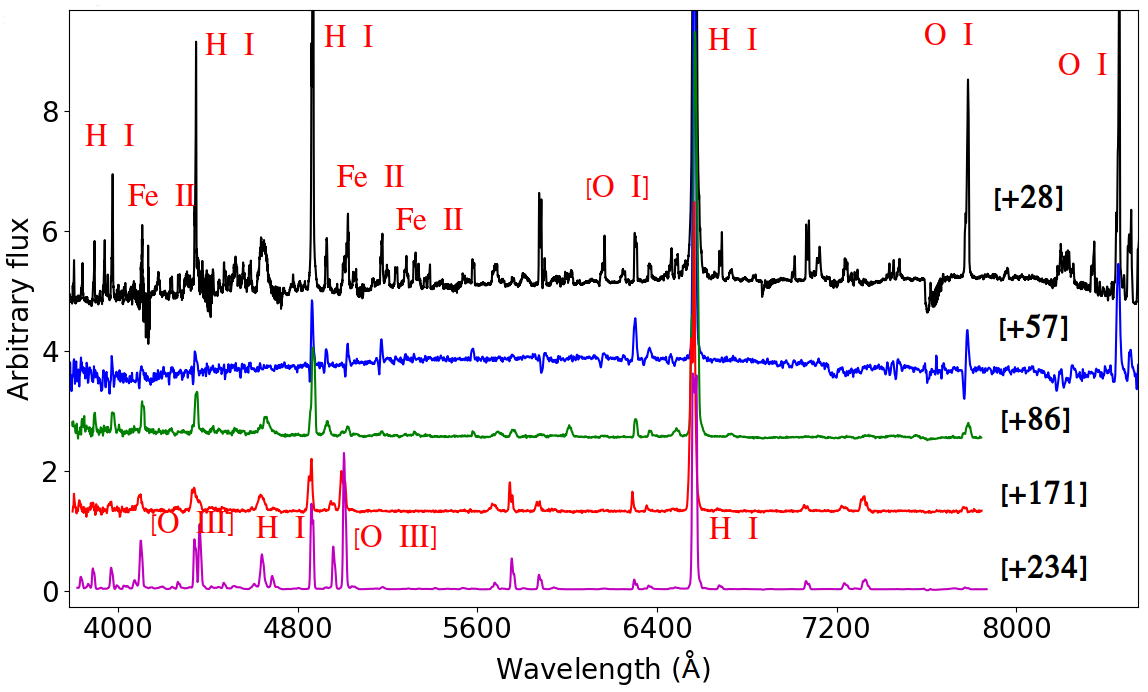}
\caption{A time sequence of the optical spectra. The numbers between brackets are days after eruption. Prominent lines are labelled; see also Table \ref{line_det}. For clarity the spectra are vertically shifted from bottom to top.}
\label{Fig:All_spec}
\end{center}
\end{figure*}

\subsection{Line identification and spectral evolution}

The first optical spectrum taken on day 10 with SOAR (Fig.~\ref{Fig:SOAR_400c2}) during the rise to the first maximum is dominated by emission lines of hydrogen and \eal{Fe}{II} with a blue continuum. The absence of P Cygni profiles is likely due to the low-resolution of the spectrum. The MIKE spectrum on day 16, which is the last spectrum before maximum light, shows similar emission lines, but relatively strong absorption components, indicating that the ejecta are optically thick in lines (e.g., \citealt{Shore_etal_2011,Williams_2013,shore_etal_2014}; see Figs.~\ref{Fig:mage_spec}). 
After maximum, the spectra (Fig.~\ref{Fig:All_spec}) were dominated by the same \eal{H}{I} and \eal{Fe}{II} emission lines accompanied with absorption components (Fig.~\ref{Fig:Hbeta}). Such spectra are of the \eal{Fe}{II} spectroscopic class \citep{Williams_2012} which typically are a characteristic of an optically thick ejecta (see e.g. \citealt{shore_etal_2014}). The spectra maintained the same emission lines all the way to the dust extinction event (day $\sim$ 125). However, the line profiles change during this time. New systems of absorption components were appearing/disappearing with maxima and minima in the optical light-curve (see Section~\ref{line_prof_sec} and Fig.~\ref{Fig:Hbeta}).

In the first two spectra taken after the dust dip on days 168 and 171, the forbidden \feal{O}{III} auroral line at 4363\,$\mathrm{\AA}$ and nebular lines at 4959 and 5007\,$\mathrm{\AA}$ appear, marking the start of the nebular phase. In Tables~\ref{line_det} and~\ref{line_det_2} we list the line identifications along with the Full Width at Half Maximum (FWHM), Equivalent Width (EW), and flux of the lines for which an estimate was possible. 

The most prominent feature in the UV spectra (Fig.~\ref{Fig:UVspectrum}) is the evidence of an overall depletion of the flux level which is probably due to absorption of the myriad of lines from singly ionized metals. Most prominent are at Day 55 the \eal{Fe}{II} UV2/UV3 absorption, while later, at Day 104, at the brink of the dust formation event, there is a prominent absorption at \eal{Fe}{II} uv1. These are all indicative of relatively high densities in the ejecta.

\subsection{Line profile evolution}
\label{line_prof_sec}
\begin{figure*}
\begin{center}
  \includegraphics[width=\textwidth]{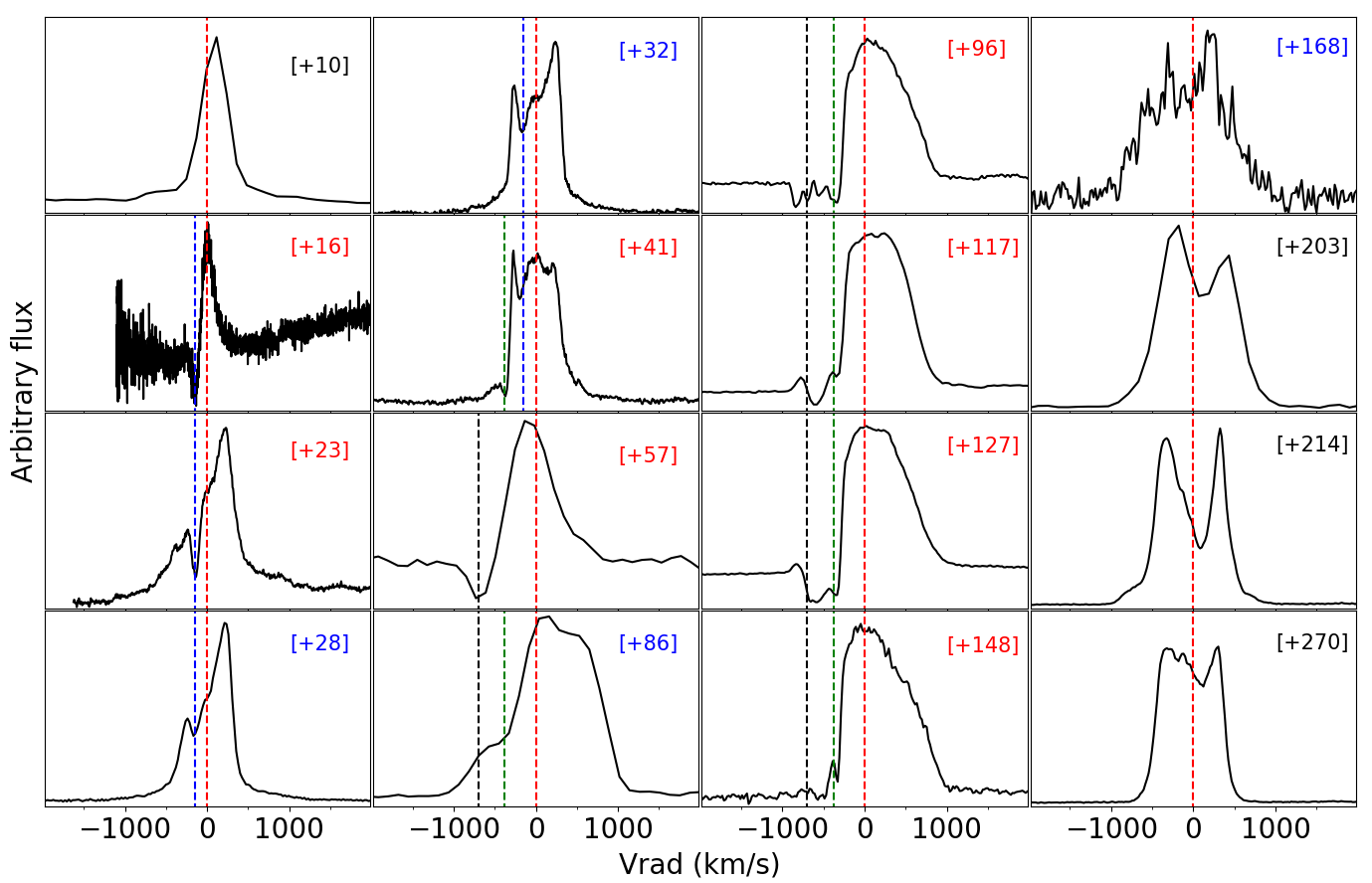}
\caption{The evolution of the H$\beta$ line profiles of ASASSN-17pf. The radial velocities are corrected to that of the LMC ($250$\,km\,s$^{-1}$) and a heliocentric correction is also applied. The red, blue, green, and black dashed lines represent $v_r$ = 0 (estimate of the line center), $-150$ (first absorption component), $-380$ (second absorption component), and $-700$\,km\,s$^{-1}$ (third absorption component), respectively. The numbers between brackets are days after discovery. For a purpose of comparison with the flares in the light-curve, the number in brackets are highlighted in red for observations taken around a maximum and in blue for these taken around a minimum. Note the line profile model overlay on day +96 in blue, which is discussed in \ref{linemodel}.}
\label{Fig:Hbeta}
\end{center}
\end{figure*}

\begin{figure}
\begin{center}
  \includegraphics[width=\columnwidth]{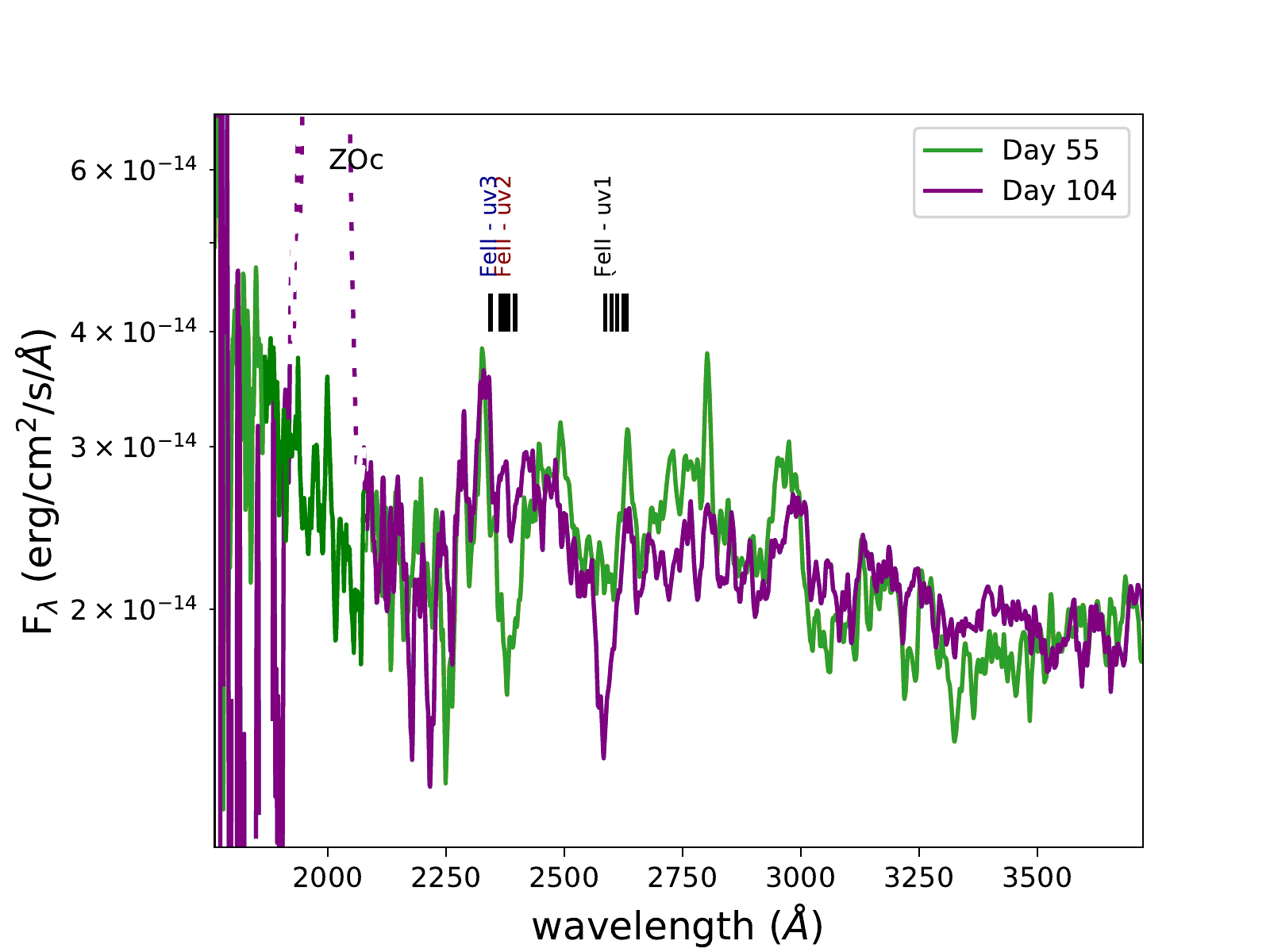}
\caption{\textit{Swift} UVOT grism spectra show absorption lines of \eal{Fe}{II} UV1 and UV2/UV3 multiplets at different dates. Contamination of a zeroth order on Day 104 has been dashed.}
\label{Fig:UVspectrum}
\end{center}
\end{figure}

\begin{figure}
\begin{center}
  \includegraphics[width=\columnwidth]{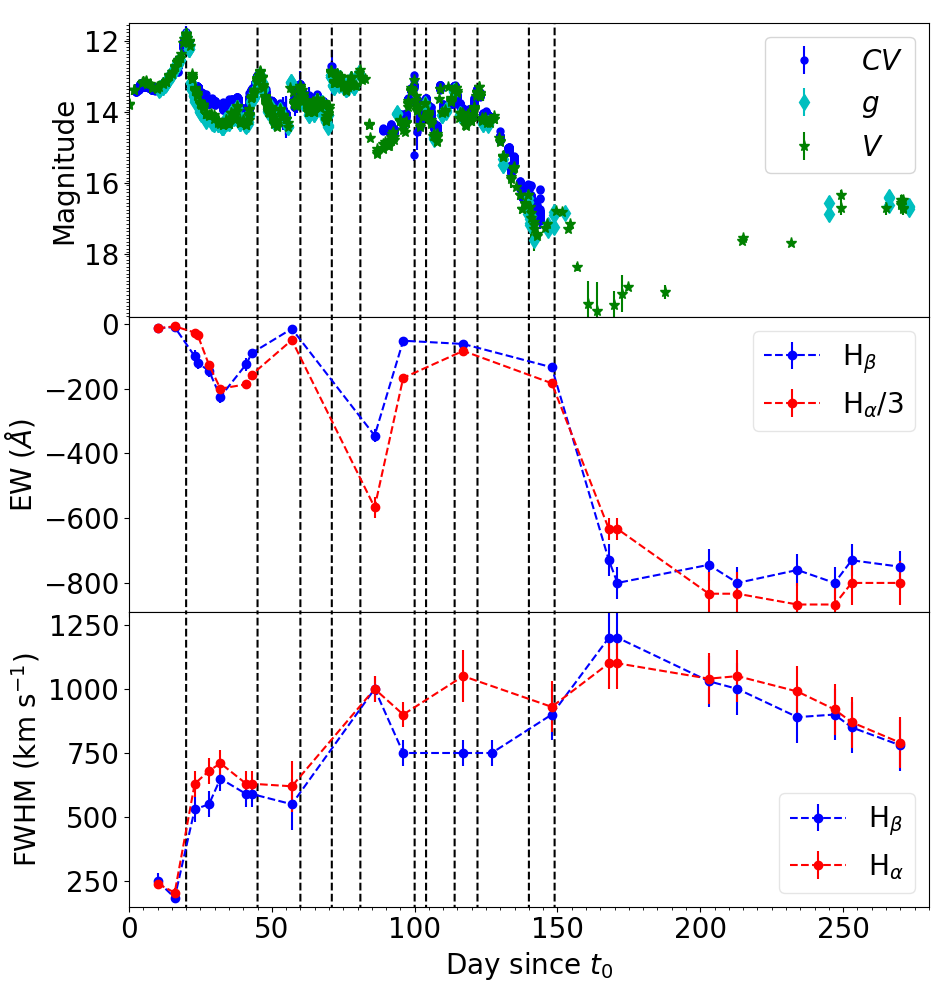}
\caption{The evolution of the EW (\textit{middle}) and FWHM (\textit{bottom}) of H$\alpha$ (red) and H$\beta$ (blue) with respect to the evolution of the optical light-curve (\textit{top}). To place the values on the same plot conveniently, we divided the EWs of H$\alpha$ by a factor of three. The black vertical dashed lines represent the dates of the flares.}
\label{Fig:LC_EW_FWHM}
\end{center}
\end{figure}

In Fig.~\ref{Fig:Hbeta} we present the evolution of the H$\beta$ line profile throughout the eruption and in Fig.~\ref{Fig:LC_EW_FWHM} we present the evolution of FWHM$_{\rm nova}$ of H$\alpha$ and H$\beta$. Measurements of the Balmer line FWHM are derived by fitting single Gaussian profiles, and correcting for the instrumental resolution: FWHM$_{\rm observed}^2$ = FWHM$_{\rm nova}^2$ + FWHM$_{\rm instrumental}^2$. 

The first spectrum taken 10 days before maximum was of relatively low resolution, and shows rounded line profiles with marginally-resolved FWHM $\approx 5.7 \pm 0.5$ \AA. Accounting for instrumental resolution, this implies a low FWHM expansion velocity of $180 \pm 30$ km\,s$^{-1}$. It is difficult to say if a weak absorption component is present at this time.
On days 16 and 23, near maximum light, an absorption component becomes prominent at a heliocentric radial velocity of $-150 \pm 10$ km\,s$^{-1}$ (marked with blue vertical lines in Fig.~\ref{Fig:Hbeta}). Near the first minimum on days 32 and 35, the absorption component almost disappeared relative to the emission lines. 

\begin{figure}
\begin{center}
  \includegraphics[width=\columnwidth]{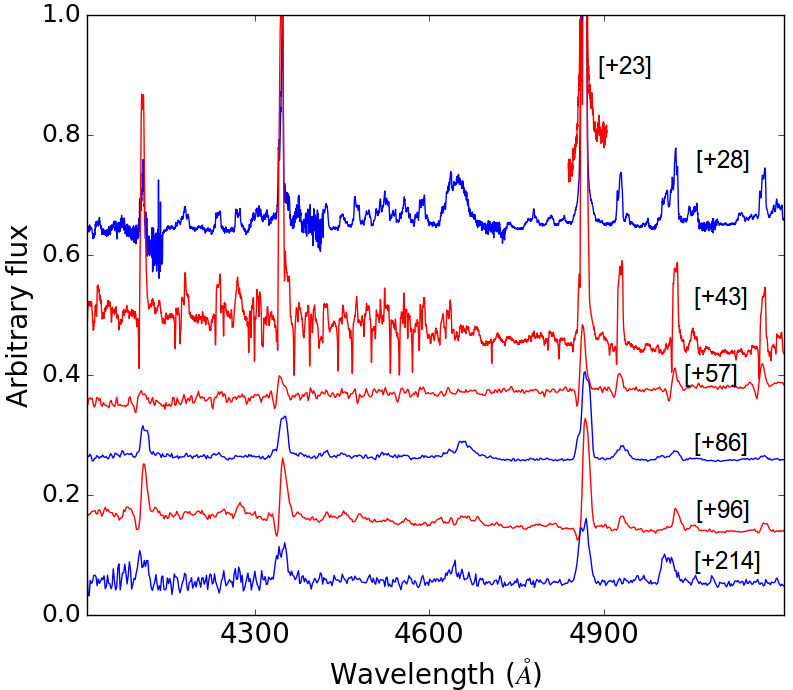}
\caption{A direct comparison between the spectra and the evolution of the P Cygni profiles. The red spectra are obtained around a maximum while the blue ones are obtained around a minimum. The numbers between brackets are days after discovery.}
\label{Fig:comp_plot}
\end{center}
\end{figure}

In the following two spectra at day 41 (same for day 43), during the rise to the second maximum, strong absorption components appear at $-380 \pm 10$ km\,s$^{-1}$ (marked with green vertical lines in Fig.~\ref{Fig:Hbeta}).
Near the third maximum on day 57, another new system of higher velocity absorption appears with velocity around $-700$ km\,s$^{-1}$ (marked with black vertical lines in Fig.~\ref{Fig:Hbeta}). These absorption components weaken around day 86 during a light-curve minimum and then reappear again on day 96 during the next flare. From day 96 and onwards, the line profiles show multiple absorption components and the light-curve shows consecutive flares which become difficult to distinguish. Fig.~\ref{Fig:comp_plot} shows how the absorption components of the P Cygni profiles strengthen around maxima and fade around minima.

In the spectra on days 117, 127, and 148, there is evidence for a decrease by around 80\,km\,$s^{-1}$ in the velocity of the second ($-380$ km\,s$^{-1}$) and third ($-700$ km\,s$^{-1}$) components (see Fig.~\ref{Fig:Hbeta}). However, some caution is required when comparing the velocities in the different observations because of the changing spectral resolution. The last spectrum showing P Cygni profiles was taken on day 148 during the dust extinction event. This coincides with the last flare in the light-curve (see Fig.~\ref{Fig:VCVVis_LC}). 

After the recovery from the dust dip, the absorption lines have completely disappeared from the spectrum and the FWHM of the lines has increased remarkably, reaching $\sim$ 1100\,km\,s$^{-1}$ (see Fig.~\ref{Fig:LC_EW_FWHM}). In the nebular phase (after day $\sim$ 200) the lines show a systematic narrowing (further discussed in Section~\ref{slow_sec}).

\subsection{Polarization}

\begin{figure}
\begin{center}
  \includegraphics[width=0.45\textwidth]{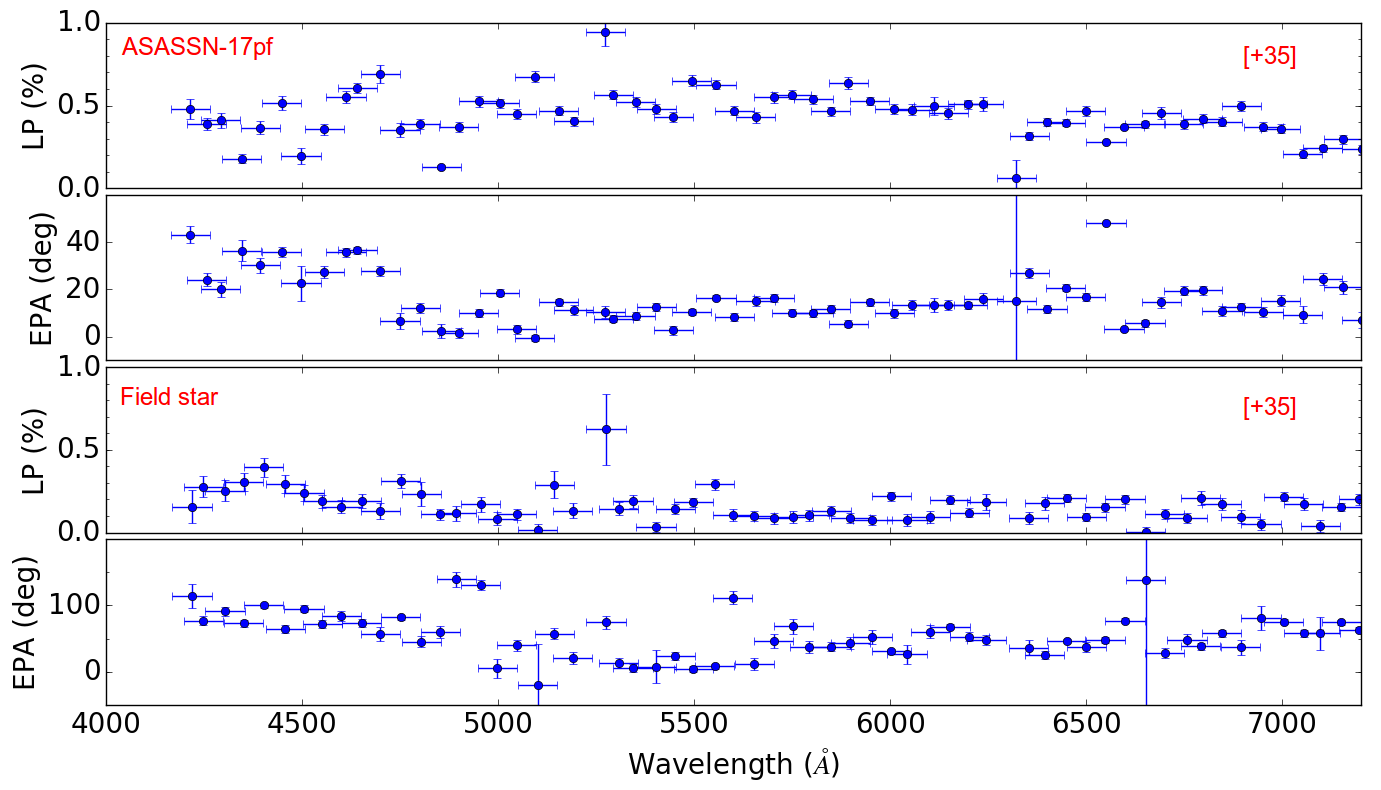}
  \includegraphics[width=0.47\textwidth]{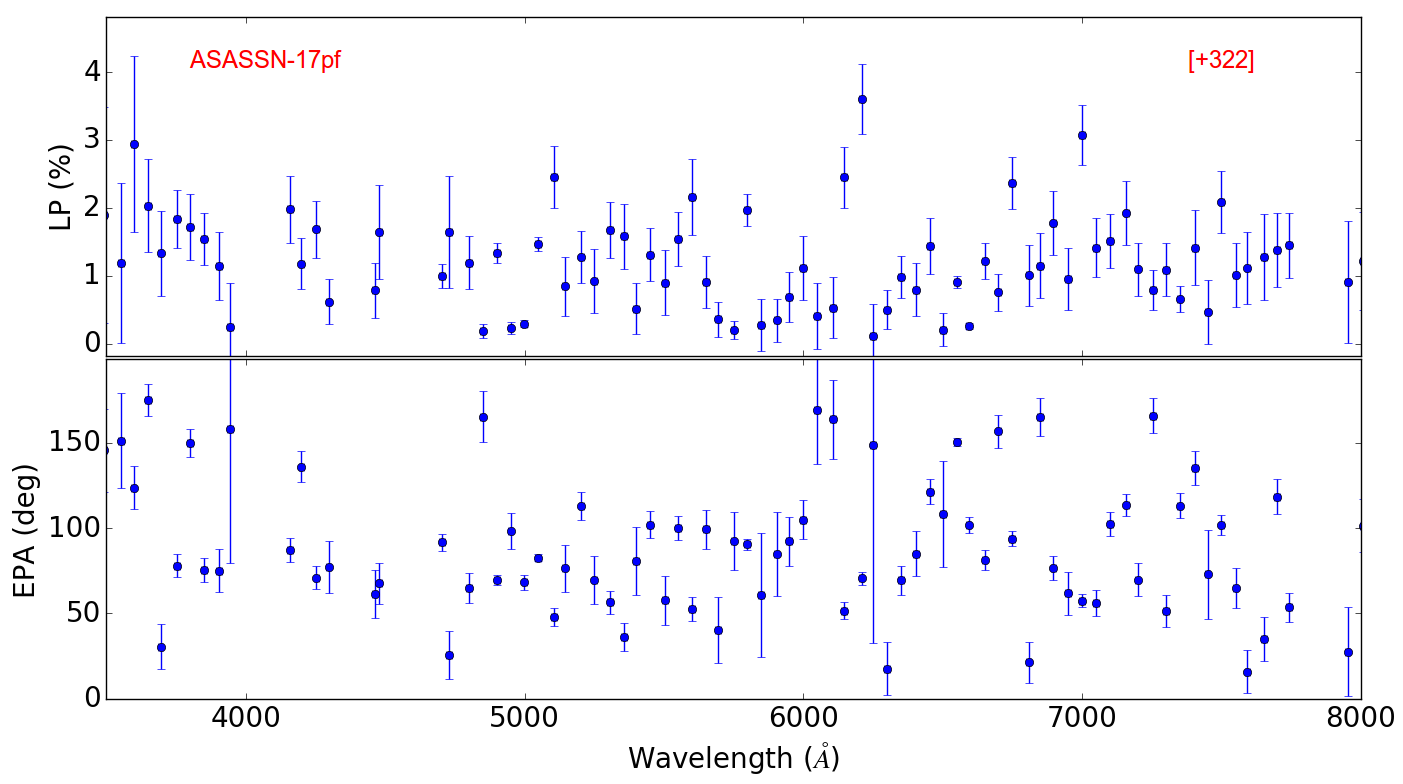}
\caption{The SALT spectropolarimetric observations of ASASSN-17pf. \textit{Top:} the linear polarization (LP) and equatorial position angle (EPA) of ASASSN-17pf in comparison with a field star taken on day 35 post-discovery. \textit{Bottom:} The LP and EPA of ASASSN-17pf taken on day 322 post-discovery. The spectral range of the second observations has been limited to 3500\,--\,8000\,$\mathrm{\AA}$ due to the large noise at the red and blue edges of the spectrum.}
\label{Fig:specpol}
\end{center}
\end{figure}

Fig.~\ref{Fig:specpol} represents the spectropolarimetric observations obtained for ASASSN-17pf on days 35 and 322. In the first observation, a field star is coincidentally in the slit and therefore we compare its spectrum with that of ASASSN-17pf to check for any intrinsic polarization from the nova apart from the interstellar polarization, assuming that the field star is in the LMC. The average linear polarization (LP) of ASASSN-17pf is ($0.45 \pm 0.04)\%$, while the field star 
shows LP ($< 0.15 \pm 0.04)\%$. The average LP of ASASSN-17pf has increased to ($1.1 \pm 0.3$)\% in the observation on day 322. The small difference between the intrinsic polarization of ASASSN-17pf and the field star on day 35 is an indication of intrinsic polarization of the nova. However, the increase in the LP of ASASSN-17pf on day 322 is strong evidence of intrinsic polarization.

\citet{Evans_etal_2002} studied a similar nova, V4362 Sgr, where they found evidence of intrinsic polarization, which was believed to have occurred prior to the dust extinction event and they concluded that the polarization is due to scattering by small dust grains, which may be the precursors of the grains that caused a later, poorly observed extinction event. However, Harvey et al. (2019 in prep.) have found that the nova may have been only detected after recovering from a dust obscuration event and it has been missed during its true peak. Therefore, the intrinsic polarization reported by \citet{Evans_etal_2002} has likely been observed after the missed dust event. For ASASSN-17pf the second spectropolarimetric observations were obtained after the recovery from the dust dip. Thus, the polarization might also be due to scattering by the remaining dust grains from the dust formation event \citep{Kucinskas_1990,Gehrz_2008}.

Another source of LP in novae is the asymmetry of the expanding ejecta (e.g., \citealt{Bjorkman_etal_1994}). In this case, one might observe a 
wavelength-dependent and time-pendent changes of position angle of polarization, which is the case by comparing both observations on days 35 and 322. \citet{Bjorkman_etal_1994} suggested that in this case, the pseudo-photosphere has became transparent to the inner-binary, which is illuminating the dispersing ejecta. If the ejecta is clumpy, it would lead to an apparent uneven illumination from the inner binary, resulting in changes in the polarization and its position angle with time.
\\
\\
\\
\\
\\
\\


\section{Discussion}
\label{sec_disc}
\subsection{Properties of ASASSN-17pf and its quiescent system}
\subsubsection{Reddening and distance}
\label{red_sec}
One of the main advantages of studying novae in the Magellanic clouds is that the distance is well-known.
For the remainder of the paper we will assume a distance $d = 50 \pm 2$ kpc to the nova \citep{Pietrzyski_etal_2013}. ASASSN-17pf's association with the LMC is consistent with the maximum magnitude of the eruption (see Section~\ref{max_mag}). It is also supported by the velocity of the absorption components of the P Cygni profiles (the radial velocity of the first absorption component would be positive without correcting the velocities to that of the LMC; see Fig.~\ref{Fig:Hbeta}).

The Galactic reddening towards the LMC is relatively low. The \citet{Schlafly_Finkbeiner_2011} recalibration of the Galactic reddening maps of \cite{Schlegel_etal_1998} indicate
$E(B-V) = 0.064$ in the direction of nova ASASSN-17pf, which should be regarded as a lower limit for any object that is sufficiently distant. As the nova is on the outskirts of the LMC, we expect little contribution from the intrinsic extinction of its host galaxy.
\\

\begin{figure}
\begin{center}
  \includegraphics[width=\columnwidth]{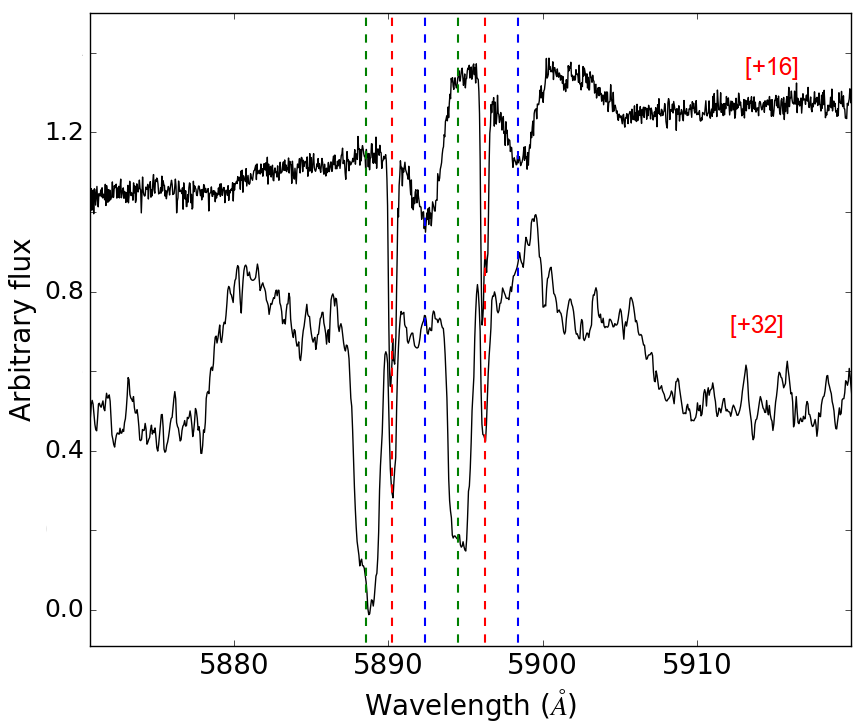}
\caption{The \eal{Na}{I}D absorption doublet at 5889.95\,$\mathrm{\AA}$ and 5895.92\,$\mathrm{\AA}$. The red dashed lines mark the narrower interstellar absorption doublet characterized by a velocity of $\sim$ 15\,km\,s$^{-1}$ (of a Galactic origin). The broader absorption components are intrinsic to the nova. On day 16, the broader absorption components are centered at a radial velocity of $\approx$ 125\,km\,s$^{-1}$ (corrected to the LMC velocity would be $\approx -125$\,km\,s$^{-1}$). On day 32, the broader absorption components are centered at a velocity of $\approx -$70\,km\,s$^{-1}$ (corrected to the LMC velocity would be $\approx -320$\,km\,s$^{-1}$). These velocities are close to the velocities of the absorption components in H$\beta$ on the same dates.}
\label{Fig:NaID}
\end{center}
\end{figure}

We use the \citet{Poznanski_etal_2012} relations to derive the extinction from the EW of the \eal{Na}{I} D1 and D2 interstellar absorption doublet at 5895.92\,$\mathrm{\AA}$ and 5889.95\,$\mathrm{\AA}$, respectively (highlighted by red vertical dashed lines in Fig.~\ref{Fig:NaID}). We differentiate between the narrow absorption components associated with interstellar absorption and the broader absorption components associated with the ejecta.
We measure EW(D1) = 0.24 $\pm$ 0.01\,$\mathrm{\AA}$ and EW(D2) = 0.33 $\pm$ 0.01\,$\mathrm{\AA}$ for the interstellar component. Thus, we derive $E(B-V) = 0.06 \pm 0.01$ and $A_V = 0.20 \pm 0.01$ for $R_V \sim 3.41$ \citep{Gordon_etal_2003}, in good agreement with the Galactic extinction maps. 

\subsubsection{Maximum magnitude and eruption amplitude}
\label{max_mag}
During the first maximum on day 20, the optical brightness of the nova reached its peak at $V_{\mathrm{max}} =$ 11.8, and correcting for dust extinction, $V_{\mathrm{0, max}} =$ 11.6. Assuming a distance of 50$\pm$2 kpc, we derive a maximum absolute magnitude $M_{\mathrm{V, max}} = -6.9 \pm 0.3$. This means that ASASSN-17pf was a relatively low-luminosity nova (see, e.g., Fig.~2 in \citealt{Shara_etal_2017_apr}). 

The nova position coincides with a source from the Optical Gravitational Lens-
ing Experiment (OGLE; \citealt{Udalski_etal_2015}) survey, which is likely the quiescent system, with a magnitude in the $I$ band of 22.7 $\pm$ 0.1 mag and a non-detection in $V$ of $>$ 22.8. Therefore, for $I_{\mathrm{max}} \simeq$ 11.6 and $V_{\mathrm{max}} \simeq$ 11.8 (Fig.~\ref{Fig:VCVVis_LC}), we derive an eruption amplitude of $\lesssim$ 11.1\,mag, which is expected for the speed class of the nova (see, e.g., the amplitude versus rate of decline diagram in \citealt{Warner_2008}).

With $I \simeq 22.73$, the absolute magnitude of the quiescent system at the distance of the LMC would be $M_I \simeq 4.2$. Based on the absolute magnitudes, periods, and spectral types of the secondary star in cataclysmic variable systems, we suggest that the secondary is possibly a K dwarf star (see, e.g., \citealt{Warner_1995} and \citealt{Darnley_etal_2012}).

\subsubsection{The slow expansion velocity and the evolution of the FWHM}
\label{slow_sec}
The spectral lines of ASASSN-17pf near the first maximum are narrow compared to other novae at these stages (see e.g. \citealt{Naito_etal_2014,Takeda_Diaz_2015}). The narrow width (FWHM $\sim$ 180\,--\,500\,km\,s$^{-1}$) of the Balmer lines (see Figs~\ref{Fig:Hbeta} and~\ref{Fig:LC_EW_FWHM}) and the low velocities of the absorption features ($\sim$ 150\,km\,s$^{-1}$) near the first maximum indicate slow expansion velocities, possibly associated with a slowly expanding dense shell or equatorial torus \citep{Williams_Mason_2010,Metzger_etal_2014,Chomiuk_etal_2014}. These are some of the lowest velocities seen in post-eruption nova spectra; similarly low velocities have only been observed in a few other slow novae, such as HR Del and V723 Cas \citep{Friedjung_1992,Goranskij_etal_2007}. \citet{Friedjung_1992} considered that the low velocities of the absorption components in HR Del pre-maximum spectra are an indication of an eruption on the surface of a low mass WD ($\sim$ 0.5\,--0.6\,M$_{\odot}$) where the conditions of the thermonuclear runaway were barely met (see as well \citealt{Kovetz_Prialnik_1985,Kato_Hachisu_1994,Iijima_etal_1998,Iijima_etal_2006}).\\


The systematic line narrowing during the nebular phase (Figs~\ref{Fig:Hbeta} and~\ref{Fig:LC_EW_FWHM}) has been observed in several novae  and attributed to the radial distribution of ejecta velocity and density (see e.g., \citealt{Shore_etal_2013,Darnley_etal_2016,Aydi_etal_2018_2,Aydi_etal_2018}).  For example, in a simple scenario with homologous ejecta (i.e., a ``Hubble flow''), the density of the outer faster-moving gas decreases rapidly compared to the inner slower-moving gas. A density decrease will lead to a decrease in the emissivity of the faster gas and to line narrowing \citep{Shore_etal_1996}.

\subsubsection{The SSS non-detection}

Novae are known to emit at constant Eddington luminosities for several days up to years after the eruption (see e.g., \citealt{Starrfield_etal_2008} and references therein). The Eddington luminosities of  0.6\,--\,1.0\,M$_{\mathrm{\odot}}$ WDs are 0.75\,--\,1.26\,$\times$\,10$^{38}$\,erg\,s$^{-1}$. If the majority of the bolometric luminosity is emitted in the soft X-rays during the supersoft source (SSS) phase, one would expect to detect supersoft emission from ASASSN-17pf using \textit{Swift}, which monitored the nova until day 357 post-eruption. Also, several LMC novae have been detected with \textit{Swift} during their SSS phases, such as nova LMC 2009a, 20012, and OGLE-2018-NOVA-01 \citep{Schwarz_etal_2015,Bode_etal_2016,ATel_11410}. Therefore, the non-detection of ASASSN-17pf with \textit{Swift}/XRT (see Section~\ref{sec_swift}) might have several interpretations:

\begin{enumerate}

\item The mass of the ejected envelope is related to the duration of $t_2$, which are both related to the SSS turn-on time. Based on the empirical relations of \citet{Henze_etal_2014_Mar}, novae with $t_2>$ 100\,d have an SSS turn-on time longer than 300 days. Thus, there is a possibility that the SSS turn-on time is longer than 357 days and the SSS phase has not started by the time of the last \textit{Swift} observation. This would also imply a low mass WD, in good agreement with the conclusions from Section~\ref{slow_sec}.

\item The SSS turned off by the time the ejecta became optically-thin to the residual hydrogen burning on the surface of the WD and therefore the SSS turn-off time occurred before the point at which the SSS would have become detectable. (see figure~8 in \citealt{Henze_etal_2014_Mar}). The duration of the SSS phase is dependent on both the WD mass and the amount of mass ejected \citep{Wolf_etal_2013}, as the SSS turn-off time is inversely proportional to the mass of the white dwarf \citep{MacDonald_1996}. Therefore, a non-detection might imply a combination of a high mass WD and a low mass remaining envelope, in contrast to the earlier conclusions. 

\item Another explanation is the presence of an additional absorber (absorbing column) to that of the Galactic interstellar medium. This additional absorber might be within the LMC and is leading to the non-detection of the supersoft X-ray emission. However, the reddening derived from the Na D lines is consistent with absorption mainly from the Galaxy (see Section~\ref{red_sec}).

\item The photospheric emission from the WD is blocked by the accretion disk \citep{Ness_etal_2013}. This might happen in the case of the system being at high inclination and if the accretion disk has survived the eruption, or the accretion has resumed soon after the eruption (see, e.g., \citealt{Walter_Battisti_2011,Aydi_etal_2018_2}). Also in this case, the supersoft X-ray spectrum would be expected to show strong emission lines \citep{Ness_etal_2013}. However, we have not obtained high-resolution optical spectroscopy for this target to confirm this suggestion.  
\end{enumerate}

\subsection{Understanding the optical flares of ASASSN-17pf}
\label{opt_flare_sec}
\subsubsection{The evolution of the P Cygni profiles}
\label{PCygni_sec}
One of the most interesting aspects of the spectral evolution of ASASSN-17pf is the variability seen in both the velocity and strength of the spectral absorption features (Figs.~\ref{Fig:Hbeta} and ~\ref{Fig:comp_plot}). These changes correspond to changes in the optical light-curve, as a new system of higher velocity absorption components appears around the same time as the appearance of a flare in the optical light-curve (Fig.~\ref{Fig:VCVVis_LC}).

Cyclic changes in the line profiles have been seen in a few other novae that also showed flares in their light-curves. \citet{Tanaka_etal_2011_159,Tanaka_etal_2011} suggested that the appearance/disappearance of an absorption component is caused by the expansion/shrinking of the optical photosphere. During a maximum, the optical photosphere expands in size and the absorption components appear as the ejecta become more optically thick. During a minimum, the optical photosphere shrinks and the absorption components weaken relative to the emission lines. 

Similar behavior was observed in novae V1369 Cen and V5668 Sgr (e.g., \citealt{2017AN....338...91J}), where new systems of absorption components at progressively higher velocities appear in the spectrum and are correlated with the appearance of flares in the light-curves. This has been attributed to multiple ejection episodes with increasing velocities (Walter et al.\ 2019, in prep.).

\begin{figure*}
\begin{center}
  \includegraphics[width=\textwidth]{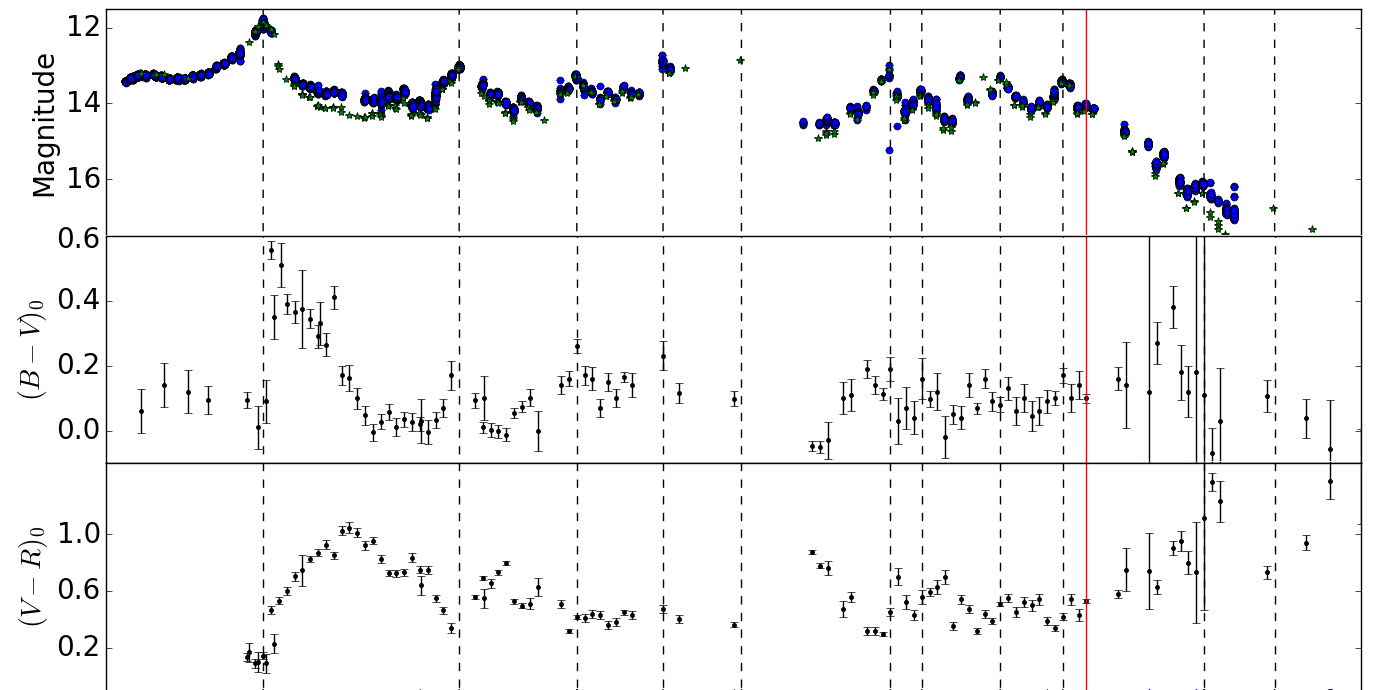}
  \includegraphics[width=\textwidth]{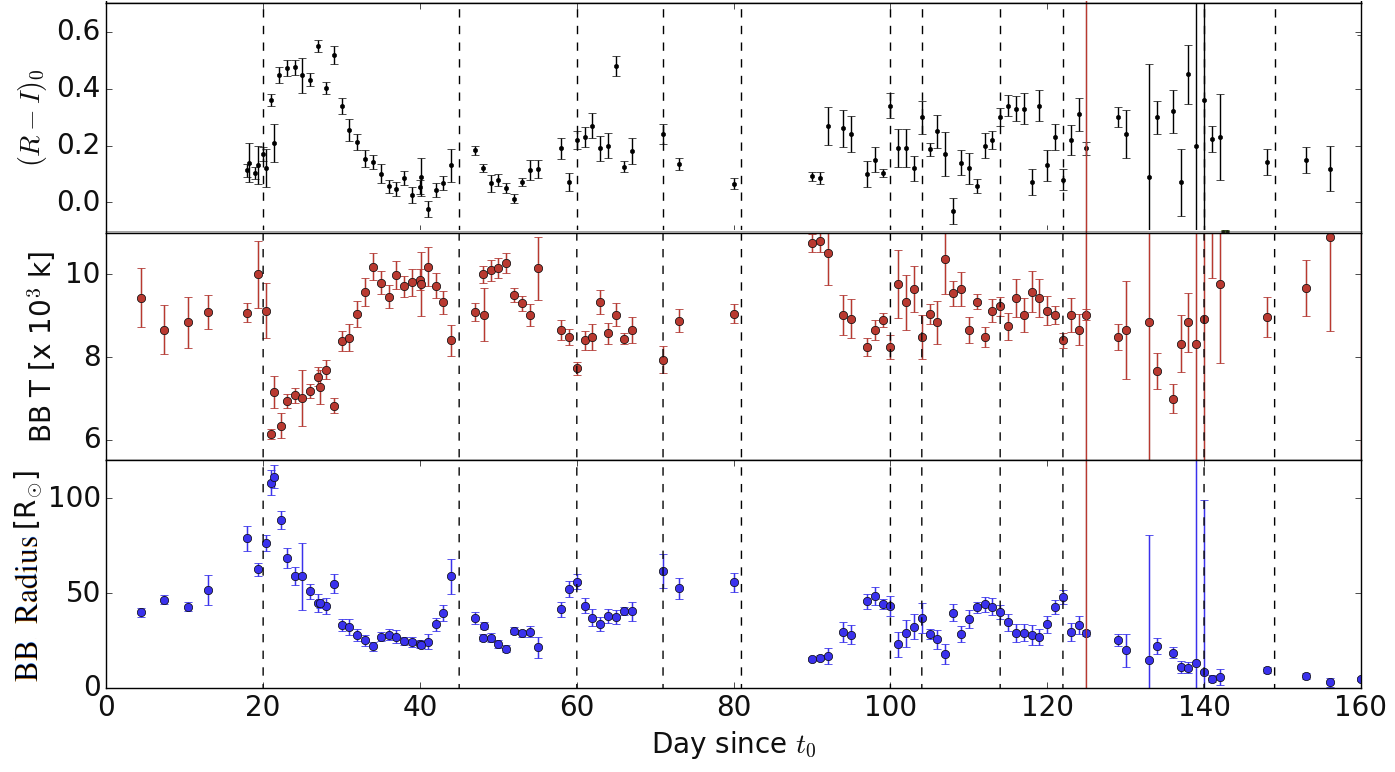}
\caption{\textit{From top to bottom}: (1) shows the $V$ and $CV$ optical light-curve of ASASSN-17pf; (2), (3), and (4) show the evolution of the optical broad-band colors $(B-V)_0$, $(V-R)_0$, and $(R-I)_0$, respectively; (5) the evolution of the blackbody (BB) temperature derived from the $(B-V)_0$ colors; and (6) the evolution of the photosphere radius. The black vertical dashed lines represent the dates of the flares. The red solid vertical line represents the estimated date of the dust dip start.}
\label{Fig:colours}
\end{center}
\end{figure*}

\subsubsection{Color evolution and photosphere modelling}
To understand more about the changing conditions during the optical flares, we consider the color evolution of ASASSN-17pf.
The evolution of the $(B - V)_0$, $(V - R)_0$, and $(R - I)_0$ colors shown in Fig.~\ref{Fig:colours} help to illustrate the changing physical conditions during the flares. The color evolution correlates with the light-curve, exhibiting significant changes coinciding with the flares. 

Nova colors are affected by the evolution of both the continuum and the emission lines. However, for ASASSN-17pf the color evolution seems to be mostly due to the evolution of the continuum. In Fig.~\ref{Fig:LC_EW_FWHM} we show the evolution of the equivalent width (EW) of the H$\alpha$ and H$\beta$ emission lines throughout the different phases of the light-curve. Remarkably, there is a trend of decreasing EW during minima and an increasing EW during maxima. This indicates that the increase in brightness during a maximum is due to an increase in the continuum flux and not the emission lines flux. The opposite is equally true during minima where the EWs of the emission lines decrease, meaning the continuum flux has decreased and the emission line flux relatively increased.


Thus, we derive the photospheric temperature and radius from the $(B-V)_0$ colors using the relations from \citet{Ballesteros_2012}. As shown in Fig.~\ref{Fig:colours}, the temperature decreases during a maximum and increases during a minimum while the radius of the photosphere expands at maximum and shrinks at minimum, consistent with our earlier interpretation. It is worth noting that while on one hand the $(R - I)_0$ colors are consistent with the $(B - V)_0$ colors, becoming bluer during a minimum and redder during a maximum, on the other hand, the $(V - R)_0$ colors show an opposite correlation. This is likely due to H$\alpha$ dominating the flux in the $R$-band, particularly during light-curve minima. Therefore, we avoid using these colors and we only make use of $(B-V)_0$ to model the photosphere.  

\begin{figure*}
\begin{center}
  \includegraphics[width=\textwidth]{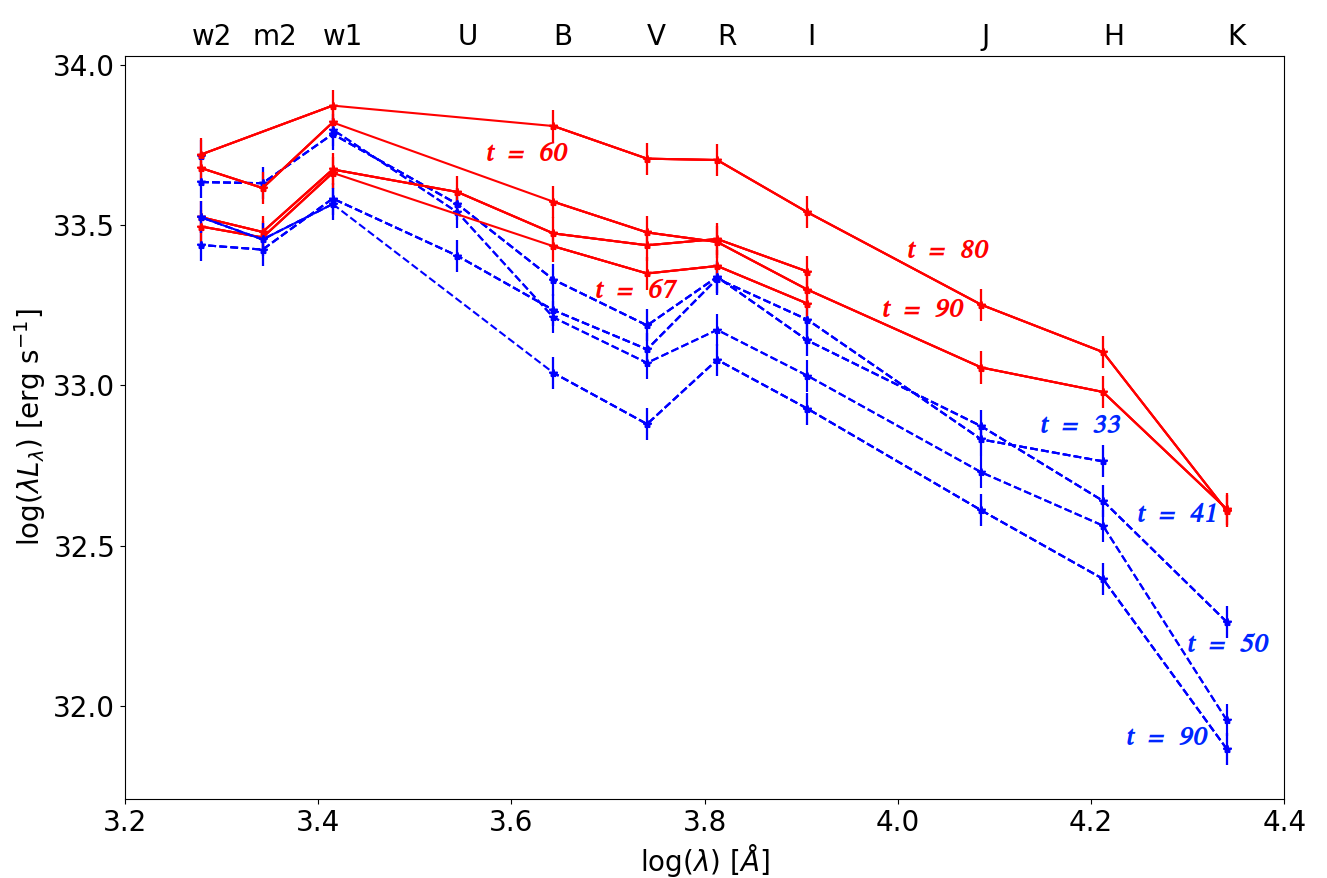}
\caption{Distance- and extinction-corrected SED plots of ASASSN-17pf. The red solid SEDs are for measurements taken near maxima, while the blue dashed SEDs are for measurements taken near minima. The error bars include contributions from the photometric and distance uncertainties. The $t-t_0$ time of the observations are also marked. 
}
\label{Fig:SEDs}
\end{center}
\end{figure*}

In order to further study the evolution of the temperature, we produce spectral energy distributions (SEDs) spanning the NUV through NIR near optical maxima (highlighted in red) and minima (highlighted in blue) in Fig.~\ref{Fig:SEDs}. Typically, during maximum optical light, the photosphere of a nova reaches its maximum radius and its effective temperature peaks in the visual light between 6000 and 8000\,K (see e.g. \citealt{Hachisu_Kato_2004,Bode_etal_2008}). \citet{Gallagher_Ney_1976} suggested that during maximum optical light, the emission can be well described by a blackbody with the wavelength of peak emission closely related to the radius of the
pseudo-photosphere. During the decline, the photosphere shrinks in radius and the emission peaks in the UV (and eventually in the X-rays) with effective temperatures $T_{\mathrm{eff}} >$ 15,000\,K (see, e.g., \citealt{Schaefer_2010, Darnley_etal_2014, Darnley_etal_2017,Aydi_etal_2018}). The SEDs of ASASSN-17pf show an oscillatory behavior: (1) during maxima the SEDs are dominated by the continuum and peak in the optical/NUV (see red lines in Fig.~\ref{Fig:SEDs}); (2) during minima the SEDs are dominated by the emission lines and show two peaks at the $R$ band (due to H$\alpha$) and at the NUV (also possibly due to emission lines; see blue lines in Fig.~\ref{Fig:SEDs}). This is additional evidence of changing photosphere radius between maxima and minima, in good agreement with Fig.~\ref{Fig:colours}, the evolution of the EW of the emission lines (Fig.~\ref{Fig:LC_EW_FWHM}), and the evolution of the P Cygni profiles (Section~\ref{PCygni_sec}). Note that the dip in the $uvm2$ (and $uvw2$) is likely due to the absorption lines in the UV spectrum (Fig.~\ref{Fig:UVspectrum}).


The rate of expansion of the photosphere is 3.5 times slower compared to the corresponding velocity of the absorption components in the optical spectra (Fig.~\ref{Fig:colours}). For example, during the rise to the second maximum, the radius of the photosphere increased by $\sim$ 40\,R$_{\odot}$ in 3 days. This coincides with the appearance of the second absorption feature with radial velocity of $-380$ km\,s$^{-1}$; during three days, the ejecta responsible for these absorption features would have traveled a distance of $\sim$ 140\,R$_{\odot}$. The discrepancy between the distances traveled by the line-absorbing ejecta and the pseudo-photosphere can have several explanations such as: (1) the radius of the pseudo-photosphere is far deeper compared to the ejecta responsible for the absorption features; (2) the material ejected at $\sim -380$\,km\,s$^{-1}$ is not a single, discrete shell that carries the pseudo-photosphere with it; (3) the color indices are contaminated by the emission lines and thus the derived radii of the pseudo-photosphere are not accurate.


\subsubsection{H$\beta$ line modelling}
\label{linemodel}
The $\textsc{shape}$ software \citep{shapenew} was used to recreate a snapshot of the spatial and velocity features of the post-eruption H$\beta$ line. $\textsc{shape}$ is a morpho-kinematical modeling tool that allows for the construction of nebular geometries in 3D that have been informed from observations. 
Resolved nova shells have been consistently described as elliptical with equatorial, tropical and polar band-like structures, and have been described as such since nova shells have been resolved (e.g. DQ Her (1934) and T Aur (1891); \citealt{Mustel_boyachuk}). 
The observed clumpy, `banded' elliptical shell structure is likely formed early since long-term observations show DQ Her and T Aur to have retained their current structure since being first resolved (see e.g. \cite{DQferland} regarding DQ Her)

\begin{figure*}
\begin{center}
  \includegraphics[width=16cm]{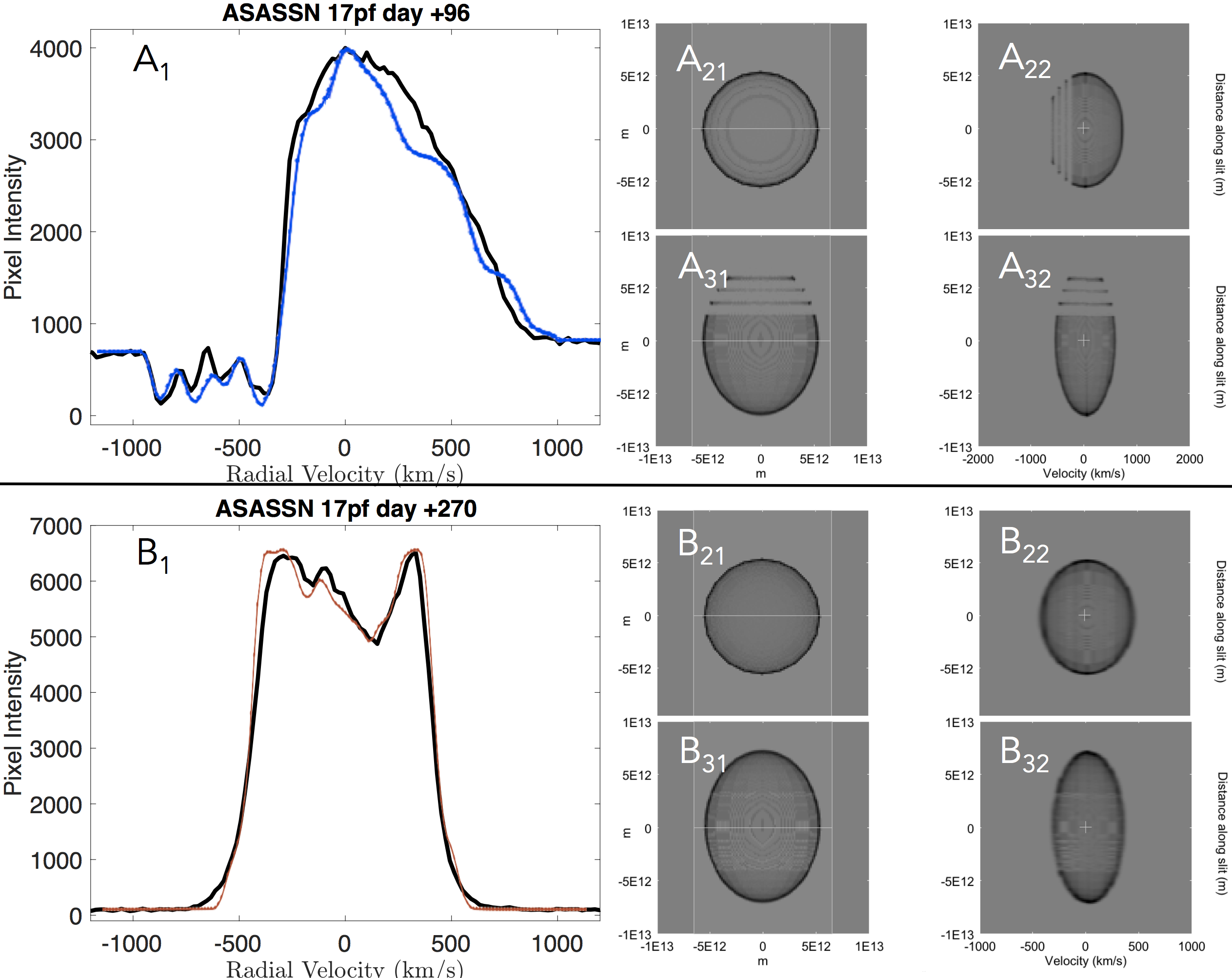}
\caption{Morpho-kinematical models produced in $\textsc{shape}$ to fit the observed H$\beta$ line profiles of two epochs. The black lines in the panels $A_{1}$ and $B_{1}$ are the H$\beta$ line profiles of days +96 ($A$) and +270 ($B$), presented earlier in Fig.~\ref{Fig:Hbeta}. The overlaid blue line (day +96, $A_{1}$) and red line (day +270, $B_{1}$) are from the $\textsc{shape}$ model fits. The same base model was used for both epochs except for the distribution and strength of the absorption bands and maximum velocity of the system. By day +270 the P Cygni profiles have disappeared and the shell emission components dominate. The lower maximum bulk velocity of day +270 is assumed to be due to the inner faster component of day +96 to have snowplowed and thus merged with the slower moving, more massive initial ejection.
Panels $A_{21}$, $A_{31}$, $B_{21}$ and $B_{31}$ are the rendered model images whereas $A_{22}$, $A_{32}$, $B_{22}$ and $B_{32}$ show the simulated position-velocity array. All $A_{x(x)}$ panels are representative of day +96 and $B_{x(x)}$ day +270. The $A_{21}$, $A_{22}$, $B_{21}$ and $B_{22}$ rendered models are viewed pole-on and $A_{31}$, $A_{32}$, $B_{31}$ and $B_{32}$ are viewed face-on (to demonstrate the model concept). If inclination was known then the distribution, covering and filling factors of the bands could be constrained. The physical size is estimated from the observed expansion velocity on day +96, although it has no effect on the final line profile fits and as such was kept constant between modelled epochs. Note the difference in velocity scales between panels $A_{22}$+$A_{32}$ versus $B_{22}$+ $B_{32}$. The continuum emission has been omitted from the 2D plots in order to clarify the simplified morpho-kinematical structure. }
\label{Fig:shape}
\end{center}
\end{figure*}

In the case of ASASSN-17pf, $\textsc{shape}$ was used to simulate the H$\beta$ line profile as observed on days +96 and +270 post-eruption (Fig.\ \ref{Fig:shape}). Based on spatially resolved imaging of $\sim$40 nova shells, basic nova shell structures have been recognized (see, e.g., \citealt{Mustel_boyachuk} and \citealt{Bode_etal_2008}) and therefore used here. 
The base morphology consists of an elliptical shell (axial ratio of 1.3) emitting in H$\beta$ and ``bands" in absorption (Fig.\ \ref{Fig:shape}). The H$\beta$ line is added as a species within the $\textsc{shape}$ model and is imported from the Kurucz spectral line database \footnote{1995 Atomic Line Data (R.L. Kurucz and B. Bell) Kurucz CD-ROM No. 23. Cambridge, Mass.: Smithsonian Astrophysical Observatory.}. The system is assumed to be in homologous expansion and viewed pole-on, with a maximum velocity of 750 km s$^{-1}$ on day +96 and 450 km s$^{-1}$ on day +270. The band distribution (see panels $A_{31}$, $B_{31}$ of Fig.\ \ref{Fig:shape}) and absorption strength were modified between epochs. On day +96 four blue-shifted narrow bands were assumed to be 100$\%$ in absorption, whereas on day +270 a wide equatorial structure in absorption for 10$\%$ of the shell thickness and 5$\%$ for a narrow blue-shifted tropical band and the rest of the shell in emission. 
As we do not have information about the system inclination, we take the pole-on assumption as the simplest case to model and to demonstrate the information that can be extracted from monitoring spectral line profile evolution post-nova. 

The elliptical shell, inner to the absorbing bands, is assigned as being the ionised region.
Absorbing bands are placed at discrete radii, as rings separated by their angles along the altitude of the shell, through the H$\beta$ emitting ellipse (with velocities dictated by the ejecta's homologous expansion, i.e. V$_{max}\times(\frac{r}{r_{0}})$).
The bands are likely precursors to an equatorial belt, tropical ring and polar cone. 
Note that the $\textsc{shape}$ model provides morpho-kinematical snapshots, but does not include hydrodynamics, which would help inform shell thickness and band structure.

The model in Fig. \ref{Fig:shape} is successful in approximating the complex P Cygni line features of ASASSN-17pf on day +96 originating from commonly observed nova shell components. The model shows an instant on day +96 when banded structure has already formed---perhaps from the interaction of several ejections. If each optical flare coincides with a new, faster ejection of material, then previously ejected material will be swept up and shocked. The distribution of these commonly observed banded structures could therefore be related to the strength and frequency of shocks in novae.

The model of day 270 shows the system to be transitioning to the nebular phase (Fig.\ \ref{Fig:shape}). 

The inner (unbroken) ionised elliptical shell can be thought of as the shell responsible for next shock, moving at a higher velocity than the post-shocked culmination of the earlier ejections.
This scenario is explored in more depth in the following section (\ref{Tyingknots}).

\subsection{Lessons from multi-wavelength synthesis}
\label{Tyingknots}

\subsubsection{Origin of the flares}
While many studies have offered explanations for the flares in nova optical light-curves, they are still poorly understood. The flares can have different timescales, lasting for a few hours/days for some novae (e.g., M31N 2008-12a, V1494 Aql, V356 Aql, ASASSN-18fv, and V1369 Cen) up to a few months for others (e.g., V723 Cas, HR Del, and V612 Sct; see \citealt{Henze_etal_2018}, \citealt{Strope_etal_2010}, and \citealt{Walter_etal_2012} \footnote{\url{http://www.astro.sunysb.edu/fwalter/SMARTS/NovaAtlas/atlas.html}}). \citet{Cassatella_etal_2004}, \citet{Csak_etal_2005}, \citet{Pejcha_2009}, and \citet{Hillman_etal_2014} have suggested that the flares are due to multiple episodes of mass ejection. However, \citet{Goranskij_etal_2007} suggested that the flares are due to instabilities in a massive accretion disk around the WD created by enhanced accretion due to nova heating of the secondary star.

Our data strongly suggest that the photosphere of ASASSN-17pf expands during the brightening of an optical flare (see as well \citealt{Tanaka_etal_2011_159} for similar conclusions). 
It appears that new, higher velocity material is ejected when a flare brightens, based on the appearance of new absorption components of increasing velocity in the spectra (see Section~\ref{line_prof_sec}). Such features have also been observed in other slow novae, such as V1369 Cen and V5668 Sgr (Walter et al.\ 2019, in prep.). Walter et al.\ suggest that the nova takes place on the surface of a low mass WD with a massive accreted envelope. Initially, the massive optically-thick envelope is partially ejected at velocities less than the escape velocity \citep[see also e.g.,][]{Starrfield_etal_1974, Sparks_etal_1978}. The nova envelope then contracts leading to optical fading and increasing densities.
This could drive enhanced nuclear burning or further ``common envelope" interaction with the binary system \citep[e.g.,][]{Livio_etal_1990,Hillman_etal_2014}. 
This in turn increases the pressure to drive further mass ejection at higher velocities. This cycle repeats until the envelope reaches escape velocity and is completely ejected.

\subsubsection{Flares-shocks link}

Growing evidence suggests that shock interactions play an important role in powering nova emission across the electromagnetic spectrum (e.g., \citealt{Weston_etal_2016,Li_etal_2017_nature}). The most striking indicator of shocks in novae is the detection of GeV $\gamma$-ray emission by \textit{Fermi}-LAT from several Galactic novae (see, e.g., \citealt{Ackermann_etal_2014,Cheung_etal_2016,Franckowiak_etal_2018}).
The $\gamma$-rays are generated by particles which are accelerated to relativistic speeds through the diffusive shock mechanism. Recent theoretical models supported by observations suggest that the shocks are formed by the interaction of two ejection components expanding at different velocities \citep{Chomiuk_etal_2014,Metzger_etal_2014,Metzger_etal_2015,Li_etal_2017_nature}. The first component is ejected early after the eruption with lower velocities and might be shaped by the binary motion into an equatorial torus. The second, faster component is ejected later with a more spherical symmetry. The faster component catches up with the slower component creating strong shocks that lead to the $\gamma$-ray emission.

The question here is whether the multiple ejections, implied by the multiple light-curve maxima and changing absorption line profiles in ASASSN-17pf, also imply formation of $\gamma$-ray emitting shocks.
Two recent novae, V1369 Cen and V5668 Sgr, have shown similar characteristics to ASASSN-17pf in their optical light-curves and spectra (Walter et al.\ 2019, in prep.). Both novae also produced $\gamma$-rays with energies $\geq$ 100\,MeV \citep{Cheung_etal_2016}. At the distance of the LMC, it is not feasible to detect $\gamma$-ray emission from ASASSN-17pf using \textit{Fermi}-LAT. Nevertheless, it seems likely that there is a link between the flares, the multiple absorption components, and nova $\gamma$-ray emission. The geometry of the multiple ejections associated with optical flares remains a mystery; it is possible that they maintain the two-component structure (equatorial torus and bipolar wind, e.g., \citealt{Chomiuk_etal_2014}) at every episode of mass ejection, or that the shock morphology is more complex as different episodes of ejection collide and merge. 

More recently, \citet{Mason_etal_2018} suggested that the GeV $\gamma$-ray emission from nova V1369 Cen is due to collision between clumps of different velocities ejected during a single ballistic ejection, contrary to the scenario of colliding winds and shells. Although this is an interesting scenario, it might not explain the increasing velocities of new absorption components appearing in the optical spectra of V1369 Cen over the first $\sim$ 50 days of eruption (note that the data interpreted by \citealt{Mason_etal_2018} were obtained at later stages).


\subsubsection{Dust formation}

About 20\% of novae show evidence of dust formation within months of the eruption \citep{Strope_etal_2010}, which manifests as a dip in the optical light-curve with associated IR excess. This only happens when the dust formed is along our line of sight. If not, an IR emission excess is seen without a dip in the optical light (e.g. N~LMC~2013, V1656~Sco, and N~Sco~2018a; see also the SMARTS atlas\footnote{\url{http://www.astro.sunysb.edu/fwalter/SMARTS/NovaAtlas/atlas.html}}). 
The mechanism responsible for dust formation is still a mystery (e.g., \citealt{Gehrz_2008}). However, recently, \citet{Derdzinski_etal_2017} suggested that strong shocks in nova ejecta might be ideal for dust formation. Derdzinski et al. proposed that dust formation occurs within the cool, dense shell behind the shock, where the density is high enough for rapid dust nucleation. 

Since ASASSN-17pf has shown a dust extinction event around 130 days post-discovery and since the very similar novae V5668 Sgr and V1369 Cen have shown evidence for dust formation, one might consider this as evidence for a link between flares, shocks, and dust. Such a link has been also suggested by \citet{Harvey_etal_2018} after a detailed study of nova V5668 Sgr. However, it is worth noting that there is no evidence for dust formation in some J-class novae (see e.g. \citealt{Strope_etal_2010}) and therefore this claim will require further investigation and detailed studies of individual novae that show flares and/or $\gamma$-ray emission. 

\section{Summary and conclusions}
\label{sec_conc}
We have presented observations of nova ASASSN-17pf, which was discovered in the LMC on HJD 2458074.7 (2017 November 17.2 UT). Optical, UV, and NIR data of this nova have led us to the following conclusions:

\begin{enumerate}
\item ASASSN-17pf is a slow nova with $t_2 >$ 100\,d, indicating a massive ejected envelope and a low-mass WD ($M_{\mathrm{WD}}<$  1.0\,M$_{\odot}$).

\item The expansion velocities indicated by the relatively narrow emission lines (FWHM $\sim$ 200\,km\,s$^{-1}$) are some of the slowest ever observed. 

\item There is a remarkable correlation between the flares in the light-curve and the emergence of new absorption features with increasing velocities in the optical spectra. This is evidence of new episodes of mass ejection associated with the flares.

\item We find another remarkable correlation between the flares and changes in the spectral energy distribution, leading us to conclude that during the flares the optical photosphere is expanding.

\item We speculate that the multiple episodes of mass ejection in novae with optical flares produce shocks which lead to high energy $\gamma$-rays and dust production.
\end{enumerate}

While we studied the condition of ASASSN-17pf during the flares, we consider that the origin of the flares is still an open question. Theoretical modelling, simulating novae in multiple dimensions, is needed to understand the physical mechanisms driving complex mass ejection in novae.
On the observational side, we note that flares appear in both slow and fast novae with different timescales, and  not all slow novae show flares nor $\gamma$-ray emission. Therefore, it is likely that there are multiple scenarios for flare formation. Hence, an observational population study of novae with flares is particularly needed. 

\section*{Acknowledgments}

We thank S. Potter and R. E. Williams for helpful discussions. We also thank F. M. Walter for help with the Chiron spectroscopy data reduction.
We acknowledge with thanks the observations used in this research from the AAVSO International Database contributed by observers worldwide, particularly Terrence Bohlsen, Nicholas Brown, Stephen Hovell, and Steve O'Connor.

EA, LC, and KVS acknowledge NSF award AST-1751874, NASA award 11-Fermi 80NSSC18K1746, and a Cottrell fellowship of the Research Corporation. JS was supported by the Packard Foundation. MJD acknowledges funding from the UK Science \& Technology Facilities Council. KLP and NPMK acknowledge support from the UK Space Agency. PC and SD acknowledge NSFC Project 11573003.  TAT acknowledges support from a Simons Foundation Fellowship  and from an IBM Einstein Fellowship from the Institute for Advanced Study, Princeton. NDR is grateful for postdoctoral support by the University of Toledo and by the Helen Luedtke Brooks Endowed Professorship. SWJ is supported by NSF award AST-1615455. DAHB gratefully acknowledges the receipt of research grants from the National Research Foundation (NRF) of South Africa.

Based on observations obtained at the Southern Astrophysical Research (SOAR) telescope, which is a joint project of the Minist\'{e}rio da Ci\^{e}ncia, Tecnologia, Inova\c{c}\~{o}es e Comunica\c{c}\~{o}es (MCTIC) do Brasil, the U.S. National Optical Astronomy Observatory (NOAO), the University of North Carolina at Chapel Hill (UNC), and Michigan State University (MSU).  This work was based in part on observations at Cerro Tololo Inter-American Observatory, National Optical Astronomy Observatory (NOAO Prop. ID: 2018A-0151; PI: N.~D.~Richardson),
which is operated by the Association of Universities for Research in Astronomy (AURA) and the SMARTS Consortium under a cooperative agreement with the National Science Foundation. A part of this work is based on observations made with the Southern African Large Telescope (SALT), with the program 2017-1-MLT-002 via Rutgers University (PI: SWJ) and the Large Science Programme on transients 2016-2-LSP-001 (PI: DAHB). 

ASAS-SN thanks the Las Cumbres Observatory and its staff for its continuing support of the ASAS-SN project. ASAS-SN is supported by the Gordon and Betty Moore Foundation through grant GBMF5490 to the Ohio State University and NSF grant AST-1515927. Development of ASAS-SN has been supported by NSF grant AST-0908816, the Mt. Cuba Astronomical Foundation, the Center for Cosmology and AstroParticle Physics at the Ohio State University, the Chinese Academy of Sciences South America Center for Astronomy (CASSACA), the Villum Foundation, and George Skestos.

\bibliography{biblio}



\appendix
 
\renewcommand\thetable{\thesection.\arabic{table}}    
\renewcommand\thefigure{\thesection.\arabic{figure}}   
\setcounter{figure}{0}

\section{Complementary plots and tables}
\label{appA}
In this Appendix we present plots and tables to complement our analysis in the main text.

\begin{figure*}[h!]
\begin{center}
  \includegraphics[width=0.8\textwidth]{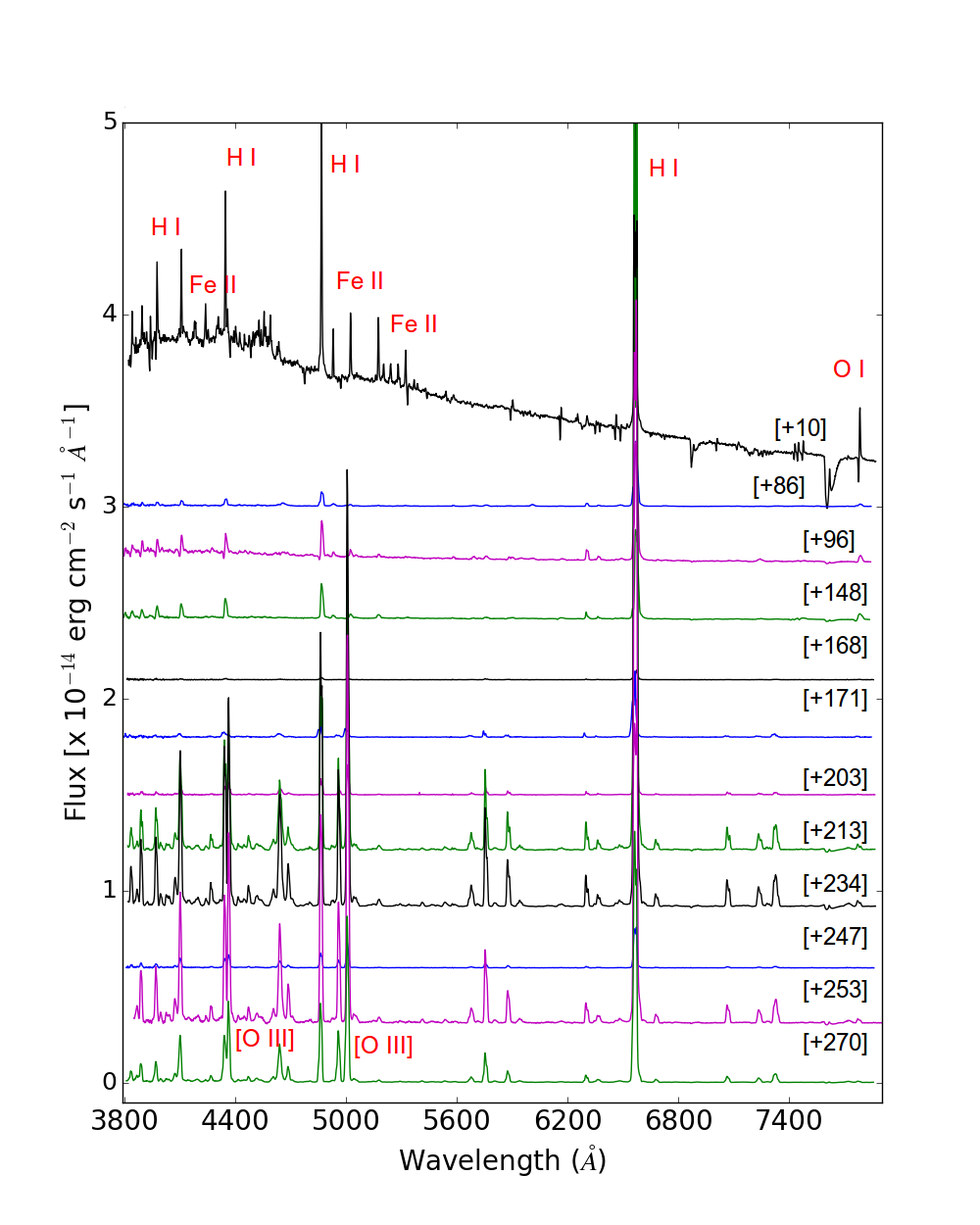}
\caption{The low-resolution SOAR flux-calibrated spectra. The numbers between brackets are days after eruption. For clarity the spectra are each vertically shifted from bottom to top by $0.3 \times 10^{-14}$ erg cm$^{-2}$ s$^{-1}$ \AA$^{-1}$.}
\label{Fig:SOAR_400c2}
\end{center}
\end{figure*}

\begin{figure*}
\begin{center}
  \includegraphics[width=0.46\textwidth]{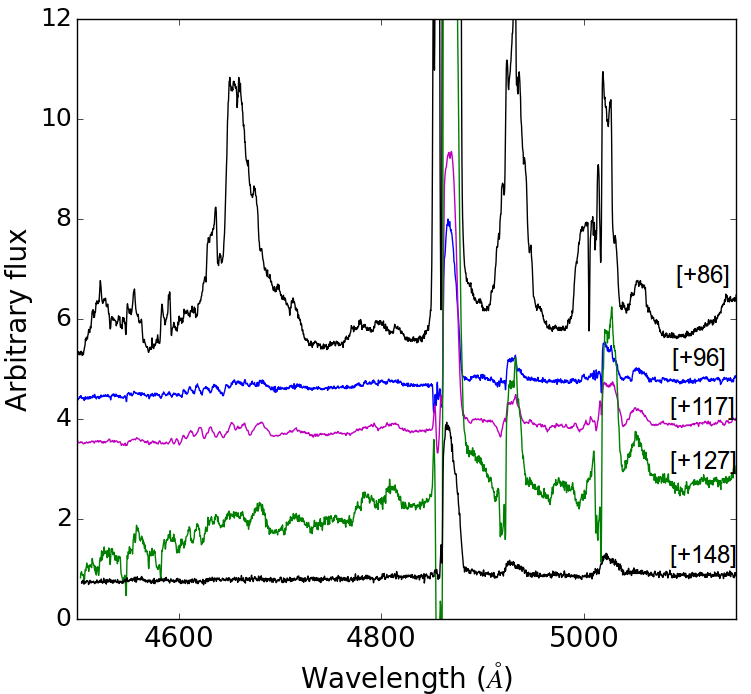}
  \includegraphics[width=0.455\textwidth]{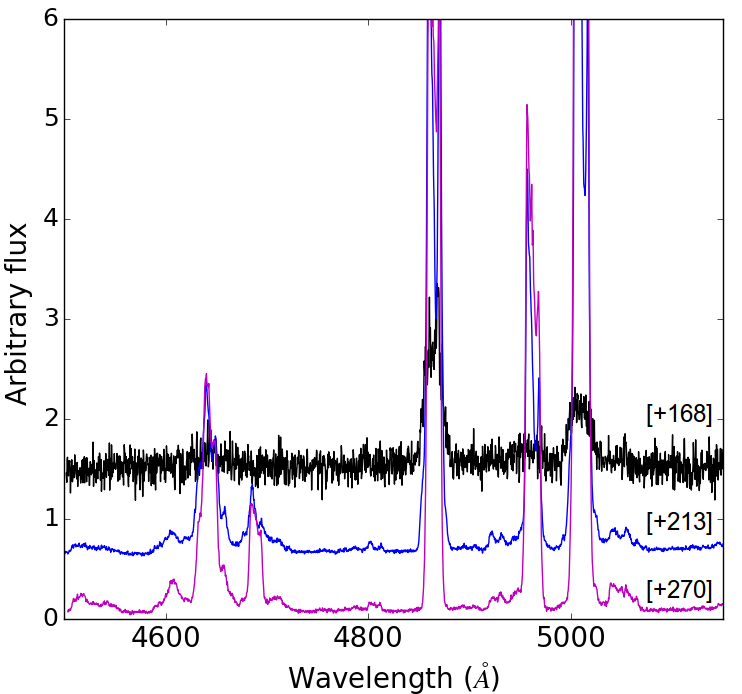}
\caption{The medium-resolution SOAR spectra. The numbers between brackets are days after eruption. For clarity the spectra are vertically shifted from bottom to top.}
\label{Fig:SOAR_2100}
\end{center}
\end{figure*}

\begin{figure}
\begin{center}
  \includegraphics[width=0.45\columnwidth]{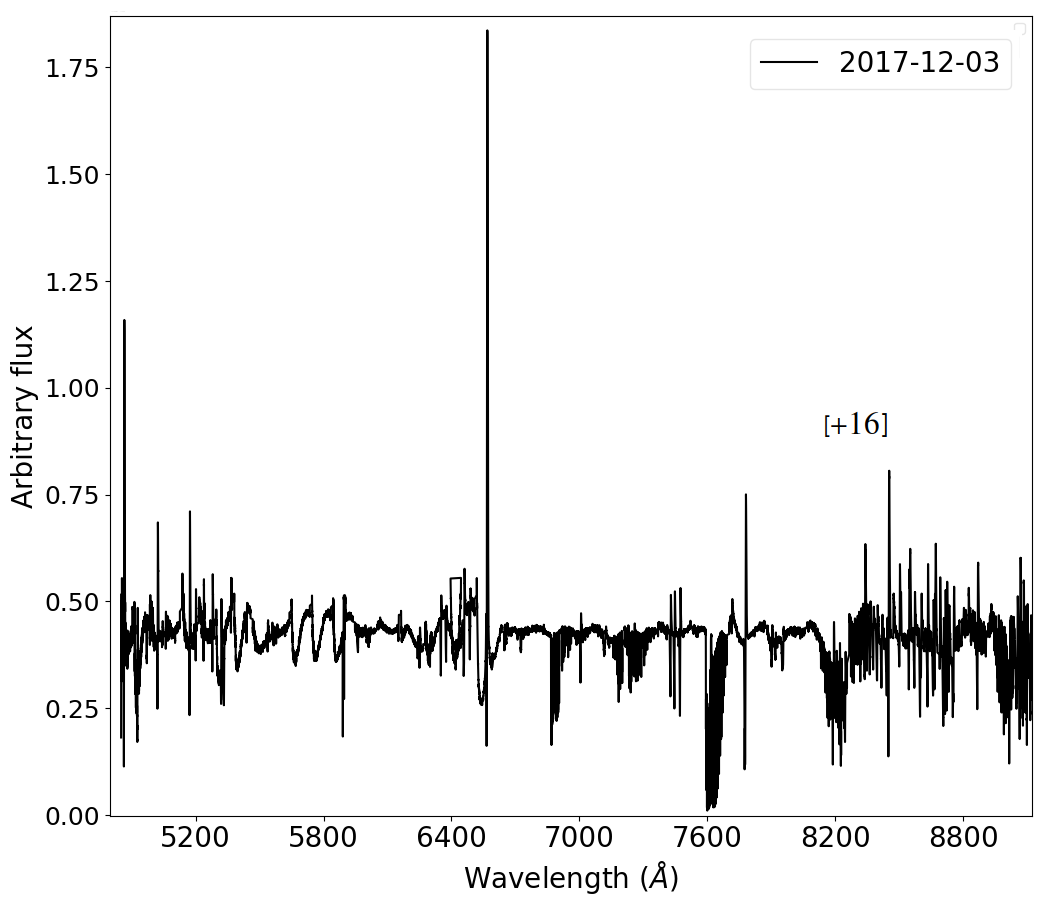}
  \includegraphics[width=0.45\columnwidth]{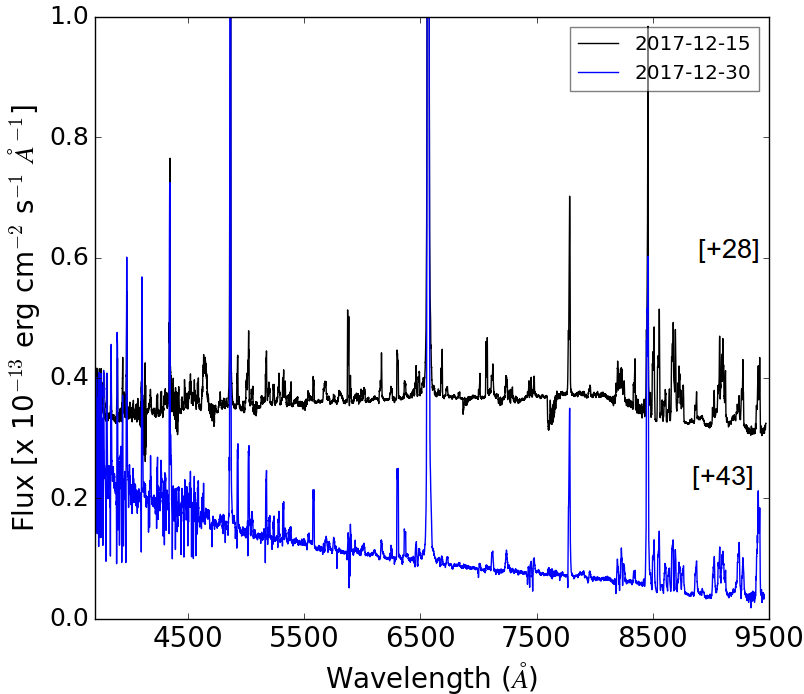}
\caption{\textit{Left:} the Magellan MIKE spectrum. \textit{Right:} Magellan MagE echelle flux-calibrated spectra. For clarity the spectrum of day 28 is vertically shifted from bottom to top by a factor of 3.0$\times10^{-14}$ erg cm$^{-2}$ s$^{-1}$ \AA$^{-1}$. The numbers between brackets are days after eruption.}
\label{Fig:mage_spec}
\end{center}
\end{figure}

\begin{figure}
\begin{center}
  \includegraphics[width=0.48\columnwidth]{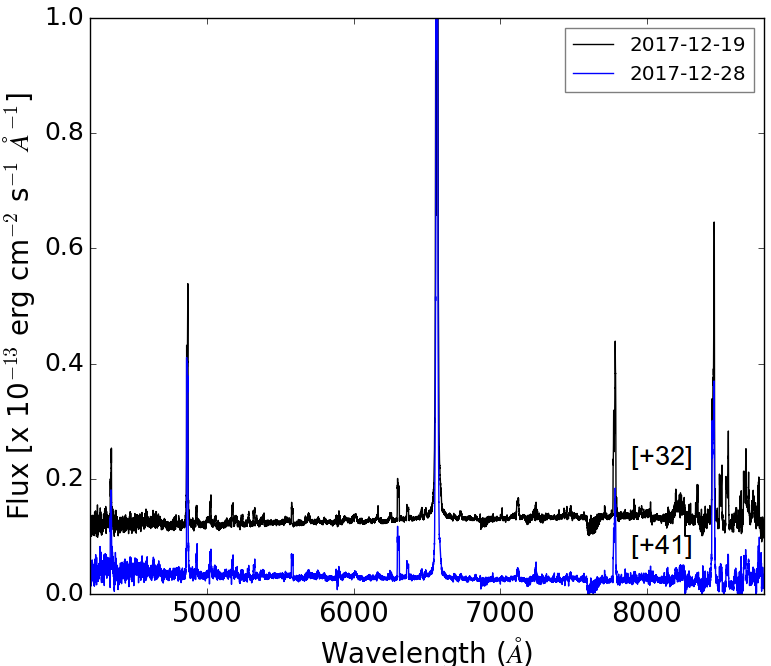}
  \includegraphics[width=0.48\columnwidth]{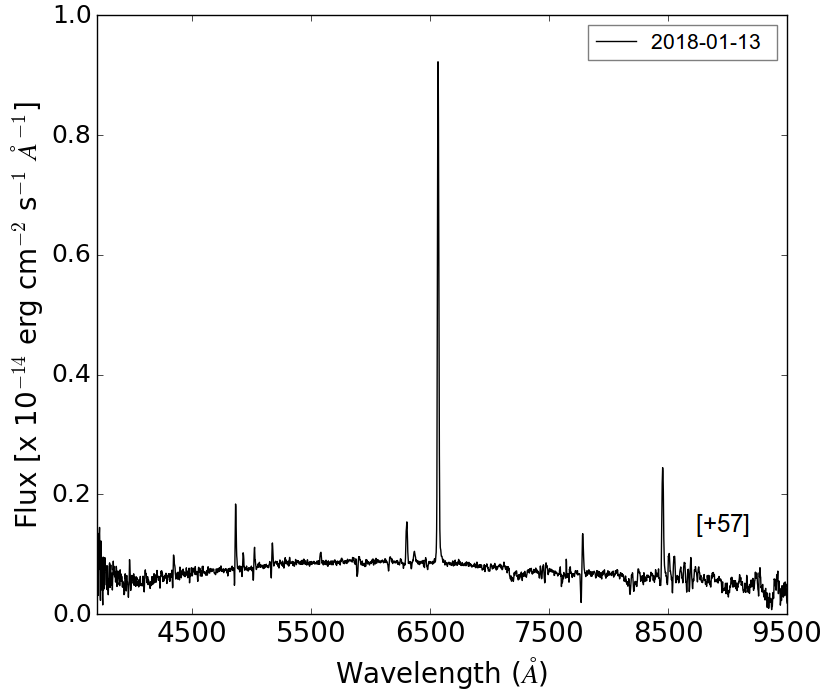}
\caption{\textit{Left:} the Chiron high-resolution, flux-calibrated echelle spectra. The numbers between brackets are days after eruption. For clarity the spectrum of day 32 is vertically shifted by a factor of 1.0$\times10^{-14}$ erg cm$^{-2}$ s$^{-1}$ \AA$^{-1}$. \textit{Right:} the Ir$\acute{\mathrm{e}}$n$\acute{\mathrm{e}}$e Du Pont WFCCD flux-calibrated spectrum of day 57.}
\label{Fig:chiron_spec}
\end{center}
\end{figure}

\begin{figure*}
\begin{center}
  \includegraphics[width=\textwidth]{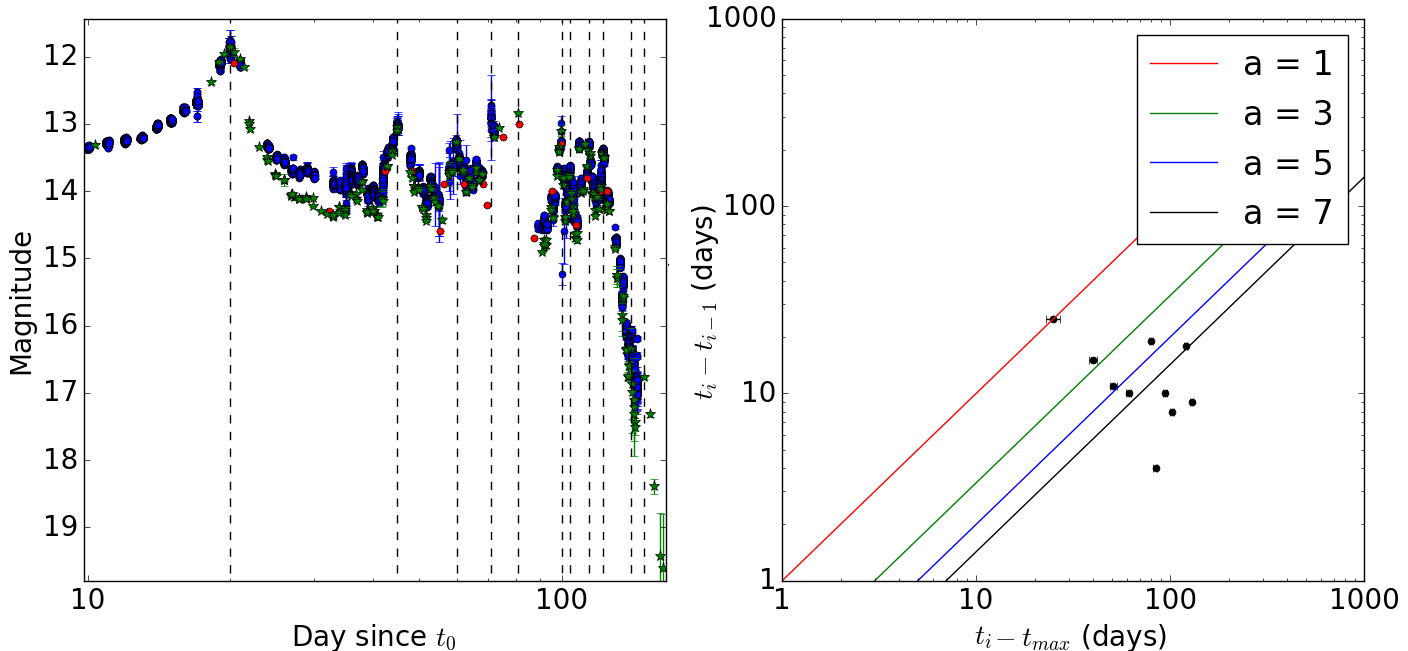}
\caption{\textit{Left:} the optical light-curve as in Fig.~\ref{Fig:VCVVis_LC} (\textit{top}) but plotted against a logarithmic scale. \textit{Right:} the time intervals between two consecutive flares $t_i - t_{i-1}$ as a function of time  elapsed since the optical maximum $t_i - t_{\mathrm{max}}$. The solid colored lines represent a power-law of the form $\log (t_i - t_{i-1}) = a + b \log (t_i - t_{\mathrm{max}})$ for different normalization factors $a$ and $b= 1$ (see \citealt{Pejcha_2009}).}
\label{Fig:time_in_Lc}
\end{center}
\end{figure*}

\begin{figure}
\begin{center}
    \includegraphics[width=0.49\columnwidth]{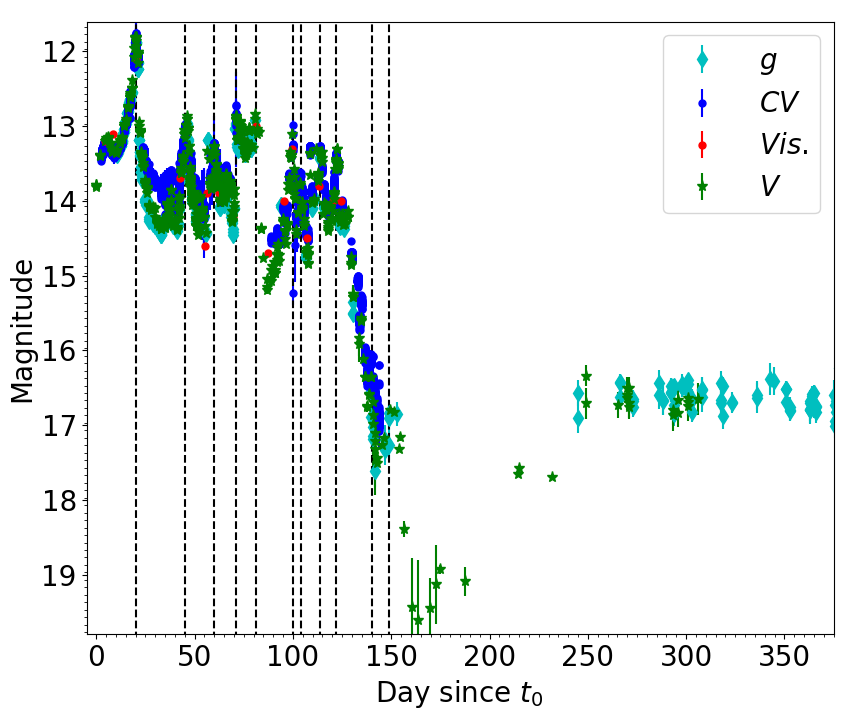}
  \includegraphics[width=0.49\columnwidth]{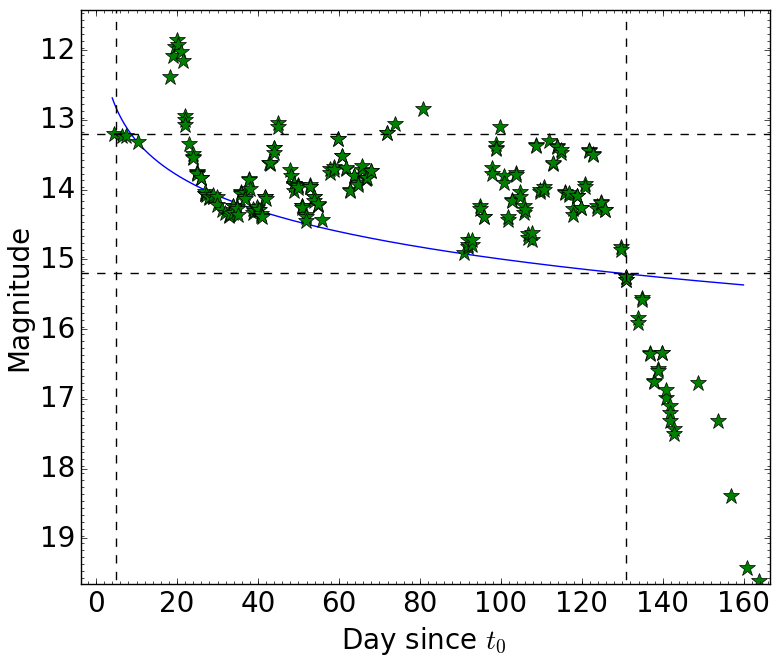}
\caption{\textit{Left:} The visual light-curve of ASASSN-17pf extending to the phase of dust recovery. \textit{Right:} the $V$-band light-curve of ASASSN-17pf. The top horizontal dashed line represents the assumed maximum peak of the base of the light-curve ($V_{\mathrm{peak, base}}$). The bottom horizontal dashed line represents $V_{\mathrm{peak, base}}$ + 2. The blue curve represents a power law fit to the base of the light-curve. The vertical dashed lines represent $t_{\mathrm{peak, base}}$ and $t_2$, respectively.}
\label{Fig:base_fit}
\end{center}
\end{figure}

\setcounter{table}{0}   
\begin{table*}
\centering
\caption{Spectral line identification of the optical spectra. We list the EW, FWHM and integrated flux of the lines for which an estimate was possible (some lines are not strong enough or blended and therefore accurate measurements for these parameters were not feasible).}
\begin{tabular}{rrrrrr}
\hline
Line & $\lambda_0$  & EW ($\lambda$) & FWHM  & Flux & ($t-t_0$)\\ 
 & ($\mathrm{\AA}$) & ($\mathrm{\AA}$)  & (km\,s$^{-1}$) & ($10^{-13}$ erg cm$^{-2}$s$^{-1}$) & (days)\\
\hline
\eal{H}{I} & 3771 & $-4.6 \pm 0.5 $ & 550 $\pm$ 30 & 1.5 $\pm 0.2$ & 43\\
\eal{H}{I} & 3798 & $-8.4 \pm 0.5 $ & 550 $\pm$ 30 & 1.9 $\pm 0.2$ & 43\\
\eal{H}{I} & 3835 & $-6.2 \pm 0.5 $ & 500 $\pm$ 30 & 1.6 $\pm 0.2$ & 43\\
\eal{H}{I} & 3889 & $-10.8 \pm 0.5 $ & 500 $\pm$ 30 & 1.6 $\pm 0.2$ & 43\\
\eal{He}{II} & 3923 & --  & -- & -- & 43\\
\eal{H}{I} & 3970 & $-8.6 \pm 0.5 $ & 430 $\pm$ 30 & 2.2 $\pm 0.2$ & 43\\
\eal{H}{I} & 4102 & $-12.1 \pm 0.5 $ & 480 $\pm$ 30 & 2.8 $\pm 0.2$ & 43\\
\eal{Fe}{II} & 4179 & - - & -- & -- & 43\\
\eal{Fe}{II} & 4233 & - - & -- & -- & 43\\
\eal{Fe}{II} & 4279 & --  & -- & -- & 43\\
\eal{Fe}{II} & 4303 & --  & -- & -- & 43\\
\eal{H}{I} & 4340 & $-36.0 \pm 2.0 $ & 550 $\pm$ 30 & 5.5 $\pm 0.2$ & 43\\
\feal{O}{III} & 4363 & -- & -- & -- & 234\\
\eal{Fe}{II} & 4352 & --  & -- & -- & 43\\
\eal{Fe}{II} & 4385 & --  & -- & -- & 43\\
\eal{Fe}{II} & 4417 & --  & -- & -- & 43\\
\eal{Fe}{II} & 4491 & --  & -- & -- & 43\\
\eal{Fe}{II} & 4508 & --  & -- & -- & 43\\
\eal{Fe}{II} & 4515 & --  & -- & -- & 43\\
\eal{Fe}{II} & 4523 & --  & -- & -- & 43\\
\eal{Fe}{II} & 4549 & --  & -- & -- & 43\\
\eal{Fe}{II} & 4556 & --  & -- & -- & 43\\
\eal{Fe}{II} & 4584 & --  & -- & -- & 43\\
\eal{N}{V} & 4609 & --  & -- & -- & 234\\
\eal{Fe}{II} & 4629 & --  & -- & -- & 43\\
\eal{N}{III} & 4638 & --  & -- & -- & 234 \\
\eal{He}{II} & 4686 & --  & -- & -- & 234\\
H$\beta$ & 4861 & $-$80 $\pm$ 5 & 590 $\pm$ 30 & 13.2 $\pm$ 2.0 & 43\\
\eal{Fe}{II} & 4924 & $-9.7 \pm 0.5 $ & 560 $\pm$ 30 & 1.5 $\pm$ 0.2 & 43\\
\feal{O}{III} & 4959 & $-253.0 \pm 20.0 $ & 770 $\pm$ 30 & 0.91 $\pm 0.1$ &234 \\
\feal{O}{III} & 5007 & $-1015.0 \pm 50.0 $ & 780 $\pm$ 30 & 3.0 $\pm 0.2$ & 234\\
\eal{Fe}{II} & 5018 & $-10.9 \pm 0.5 $ & 570 $\pm$ 30 & 1.6 $\pm  0.2$ & 43\\
\eal{He}{I} & 5048 & $-3.1 \pm 0.5 $ & -- & -- & 43\\
\eal{Fe}{II} & 5169 & $-8.6 \pm 0.5 $ & 530 $\pm$ 30 & 1.7 $\pm 0.2$ & 43\\
\eal{Fe}{II} & 5198 & -- & -- & -- & 43\\
\eal{Fe}{II} & 5214 & -- & -- & -- & 43\\
\eal{Fe}{II} & 5235 & -- & -- & -- & 43\\
\eal{Fe}{II} & 5276 & -- & -- & -- & 43\\
\eal{Fe}{II} & 5317 & -- & -- & -- & 43\\
\eal{Fe}{II} & 5363 & -- & -- & -- & 43\\
\eal{He}{II} & 5412 & -- & -- & -- & 234\\
\eal{Fe}{II} & 5425 & -- & -- & -- & 43\\

\hline
\end{tabular}
\label{line_det}
\end{table*}

\begin{table*}
\centering
\caption{Table~\ref{line_det} continued.}
\begin{tabular}{rrrrrr}
\hline
Line & $\lambda_0$  & EW ($\lambda$) & FWHM  & Flux & ($t-t_0$)\\ 
 & ($\mathrm{\AA}$) & ($\mathrm{\AA}$)  & (km\,s$^{-1}$) & ($10^{-13}$ erg cm$^{-2}$s$^{-1}$) & (days)\\
\hline
\eal{Fe}{II} & 5535 & -- & -- & -- & 43\\
\feal{O}{I} & 5577  & $-8.9 \pm 0.5 $ & 590 $\pm$ 30 & 1.1 $\pm 0.2$ & 43\\
\eal{Na}{I} & 5686 & -- & -- & -- & 43\\
\eal{Al}{II} & 5706 & -- & -- & -- & 43\\
\feal{N}{II} & 5755 & -- & -- & -- & 43\\
\eal{Na}{I} & 5892 & -- & -- & -- & 43\\
\eal{Na}{I} & 6159 & -- & -- & -- & 43\\
\eal{Fe}{II} & 6248 & -- & -- & -- & 43\\
\feal{O}{I} & 6300  & $-19.0 \pm 0.5 $ & 570$\pm$ 30 & 2.1 $\pm 0.2$ & 43\\
\feal{O}{I} & 6364 & -- & -- & -- & 43\\
\eal{Fe}{II} & 6456 & -- & -- & -- & 43\\
\eal{N}{I} & 6486 & -- & -- & -- & 43\\
H$\alpha$ & 6563 & $-$345$\pm$ 50 & 630 $\pm$ 30 & 27.3 $\pm$ 2.0 & 43\\
\eal{He}{II} & 6560 & --  & -- & -- & 43\\
\eal{He}{I} & 6687 & --  & -- & -- & 43\\
\feal{S}{II} & 6731 ? & --  & -- & -- & 43\\
?? & 7015  & --  & -- & -- & 43\\
\eal{He}{I} & 7065 & --  & -- & -- & 43\\
?? & 7118  & --  & -- & -- & 43\\
\feal{C}{II} & 7235 ? & --  & -- & -- & 43\\
\feal{O}{II} & 7320/30 & -- & -- & -- & 234\\
?? & 7431  & --  & -- & -- & 43\\
?? & 7451  & --  & -- & -- & 43\\
\eal{Al}{II} & 7477 ? & --  & -- & -- & 43\\
\eal{O}{I} & 7773  & $-50.0 \pm 2.0 $ & 500$\pm$ 30 & 3.7 $\pm 0.2$ & 43\\
?? & 7958  & --  & -- & -- & 43\\
\eal{Na}{I} & 8191 & --  & -- & -- & 43\\
\eal{Ca}{II} & 8251 & --  & -- & -- & 43\\
\eal{C}{I} & 8335 & --  & -- & -- & 43\\
\eal{O}{I} & 8446 & $-188.0 \pm 10.0$ & 580$\pm$ 30 & 10.0 $\pm$ 1.0 & 43\\
\eal{H}{I} & 8502 & -- & -- & -- & 43\\
\eal{H}{I} & 8545 & -- & -- & -- & 43\\
\eal{H}{I} & 8595 & -- & -- & -- & 43\\
\eal{H}{I} & 8665 & -- & -- & -- & 43\\
\eal{H}{I} & 8750 & -- & -- & -- & 43\\
\eal{H}{I} & 8863 & -- & -- & -- & 43\\
\eal{H}{I} & 9015 & -- & -- & -- & 43\\
\eal{N}{I} & 9060 & -- & -- & -- & 43\\
\eal{C}{I} & 9087 & -- & -- & -- & 43\\
\eal{H}{I} & 9229 & -- & -- & -- & 43\\
\eal{O}{I} & 9264 & -- & -- & -- & 43\\
\eal{N}{I} & 9395 & -- & -- & -- & 43\\
\eal{C}{I} & 9406 & -- & -- & -- & 43\\
\hline
\end{tabular}
\label{line_det_2}
\end{table*}

\clearpage

\section{Logs of the observations}
\label{appB}
In this Appendix we present logs of the observations.

\setcounter{table}{0}   
\begin{table}[h!]
\centering
\caption{A sample of the SMARTS \textit{BVRIJHK} photometry. The time series photometry is available on the electronic version.}
\begin{tabular}{rrrrr}
\hline
HJD & $(t - t_0)$   & Band & Magnitude & Magnitude Err.\\ 
& (days) & & (mag) & (mag)\\
\hline
2458093.722 & 19   & B &   12.246 &   0.013\\
2458093.725 & 19   & V &   12.081 &  0.020\\
2458093.727 & 19   & R &   11.905 &   0.019\\
2458093.730 & 19   & I   & 11.742 &  0.013\\
2458093.722 & 19   & H &   10.880 &   0.559\\
2458093.723 & 19  & J  &  11.350 &   0.296\\
2458093.724 & 19  & K &   11.477 &   0.675\\
2458094.710 & 20  & V &   11.862 &   0.027\\
2458094.712 & 20   & R &   11.730 &   0.016\\
2458094.714 & 20   & I   & 11.577  &  0.012\\
2458094.710 & 20   & H &   10.751 &   0.619\\
2458094.711 & 20   & J  &  11.196 &   0.178\\
2458094.711 & 20   & K &   11.318 &   0.769\\
\hline
\end{tabular}
\label{table:SMARTS}
\end{table}

\begin{table}[h!]
\centering
\caption{A sample of the broadband photometery obtained at the Kleinkaroo Observatory. The time series photometry is available on the electronic version.}
\begin{tabular}{rrrrrrrr}
\hline
HJD	& $(t-t_0)$ & CCD-$V$ &	$V$ Err. & CCD-$Ic$ & $Ic$ Err.& CCD-$Rc$ & $Rc$ Err.\\
& (days) & (mag) & (mag) & (mag) & (mag) & (mag) & (mag)\\
\hline
2458095.32 & 20.62 & 11.91 & 0.02 & 11.57 & 0.02 & 11.73 & 0.02\\
2458095.41 & 20.72 & 11.91 & 0.02 & 11.57 & 0.02 & --  & --\\	 
2458095.50 & 20.81 & 11.94 & 0.02 & 11.58 & 0.02 & 11.73 & 0.02\\
2458098.51 & 23.82 & 13.49 & 0.04 & 12.32 & 0.03 & 12.89 & 0.03\\
2458102.36 & 27.67 & 14.12 & 0.05 & 12.70 & 0.03 & 13.24 & 0.04\\
2458102.43 & 27.74 & 14.13 & 0.05 & 12.72 & 0.03 & 13.25 & 0.04\\
2458102.53 & 27.84 & 14.11 & 0.05 & 12.67 & 0.03 & 13.23 & 0.04\\
2458103.53 & 28.84 & 14.08 & 0.05 & 12.68 & 0.03 & 13.20 & 0.04\\
2458107.36 & 31.67 & 14.43 & 0.05 & 13.07 & 0.03 & 13.41 & 0.04\\
2458107.42 & 31.73 & 14.38 & 0.05 & 13.03 & 0.03 & 13.39 & 0.04\\
2458108.33 & 32.64 & 14.22 & 0.05 & 13.05 & 0.03 & 13.33 & 0.04\\
2458108.37 & 32.68 & 14.21 & 0.05 & 13.04 & 0.03 & 13.34 & 0.04\\
2458113.36 & 37.67 & 14.23 & 0.05 & 13.40 & 0.04 & 13.55 & 0.04\\
2458113.42 & 37.73 & 14.24 & 0.05 & 13.38 & 0.04 & 13.55 & 0.04\\
2458113.51 & 37.82 & 14.35 & 0.05 & 13.39 & 0.04 & 13.62 & 0.04\\
\hline
\end{tabular}
\label{table:berto}
\end{table}

\begin{table}
\centering
\caption{\emph{Swift} observations of ASASSN-17pf used in this work.}
\begin{tabular}{cccc}
\hline
Date & t - t$_0$ & Obs ID & Exposure\\
     &   (days)  &        &  (s)\\
\hline
2017-12-01 UT03:29:56 & 13 & 10446001 & 2033 \\
2017-12-21 UT00:28:57 & 33 & 10446002 & 466 \\
2017-12-29 UT13:54:57 & 42 & 10446003 & 1437 \\
2018-01-05 UT01:59:57 & 48 & 10446004 & 983 \\
2018-01-11 UT07:48:57 & 55 & 10511001 & 267 \\
2018-01-11 UT07:53:57 & 55 & 10511002 & 2954 \\
2018-01-11 UT12:31:57 & 55 & 10511003 & 250 \\
2018-01-11 UT12:36:57 & 55 & 10511004 & 2534\\
2018-01-17 UT16:56:57 & 61 & 10446006 & 456 \\
2018-01-24 UT03:26:57 & 67 & 10446007 & 846 \\
2018-02-04 UT18:59:57 & 79 & 10446008 & 978 \\
2018-02-06 UT17:10:57 & 81 & 10551002 & 4952 \\
2018-02-07 UT12:22:57 & 82 & 10446009 & 993 \\
2018-02-10 UT19:49:57 & 85 & 10446010 & 938 \\
2018-02-14 UT19:24:57 & 89 & 10446011 & 218 \\
2018-02-16 UT06:23:57 & 91 & 10446012 & 925 \\
2018-02-16 UT11:14:57 & 91 & 10511006 & 4529 \\
2018-02-19 UT17:21:57 & 94 & 10446013 & 1073 \\
2018-02-22 UT17:09:56 & 97 & 10446014 & 1098 \\
2018-02-25 UT05:28:57 & 100 & 10446015 & 1083 \\
2018-02-28 UT01:58:56 & 102 & 10446016 & 1053 \\
2018-03-01 UT05:24:57 & 104 & 10511008 & 5792 \\
2018-03-04 UT09:45:57 & 107 & 10446017 & 1008 \\
2018-03-07 UT06:15:57 & 110 & 10446018 & 1028 \\
2018-03-10 UT01:07:57 & 112 & 10446019 & 1066 \\
2018-03-13 UT23:01:57 & 116 & 10446020 & 963 \\
2018-03-28 UT19:10:57 & 131 & 10446021 & 968 \\
2018-04-01 UT15:22:57 & 135 & 10446022 & 406 \\
2018-04-10 UT23:58:57 & 144 & 10446023 & 812 \\
2018-04-15 UT15:48:57 & 149 & 10446024 & 860 \\
2018-04-22 UT22:43:57 & 156 & 10446025 & 1251 \\
2018-04-29 UT01:14:56 & 162 & 10446026 & 1116 \\
2018-05-18 UT11:20:57 & 182 & 10446027 & 5578 \\
2018-05-20 UT14:15:57 & 184 & 10446028 & 988 \\
2018-06-02 UT01:48:57 & 196 & 10446029 & 4102 \\
2018-07-07 UT08:32:57 & 232 & 10446030 & 562 \\
2018-08-03 UT00:33:57 & 258 & 10446031 & 1449 \\
2018-08-18 UT00:45:56 & 273 & 10446032 & 653 \\
2018-08-31 UT19:16:56 & 287 & 10446033 & 1550 \\
2018-09-14 UT05:08:56 & 301 & 10446034 & 1399 \\
2018-09-28 UT09:57:56 & 315 & 10446035 & 1168 \\
2018-10-26 UT12:22:57 & 343 & 10446036 & 1427 \\
2018-11-09 UT01:34:56 & 356 & 10446037 & 762 \\
\hline
\end{tabular}
\label{table:swift}
\end{table}

\begin{sidewaystable}[h!]
\centering
\caption{A sample of the \textit{Swift} UVOT photometry. The time series photometry is available on the electronic version.}
\begin{tabular}{rrrrrrrrrrrrrrr}
\hline
HJD & $(t-t_0)$ & $V$ & $V$ Err. & $B$ & $B$ Err. & $U$ & $U$ Err. & $UVW1$ & $UVW1$ Err. & $UVM2$ & $UVM2$ Err. & $UVW2$ & $UVW2$ Err.\\
& (days) & \multicolumn{12}{c}{(mag)}\\
\hline
2458088.3 & 13.6 & 13.00 & 0.03 & 13.16 & 0.02 & 12.34 & 0.03 & 12.86 & 0.03 & 13.24 & 0.03 & 13.30 & 0.03\\
2458108.0 & 33.3 & 14.31 & 0.05 & 14.33 & 0.03 & 13.21 & 0.03 & 13.10 & 0.03 & 13.17 & 0.03 & 13.32 & 0.03\\
2458116.6 & 41.9 & 13.77 & 0.03 & 13.82 & 0.02 & 12.81 & 0.03 & 12.60 & 0.02 & 12.66 & 0.03 & 12.80 & 0.02\\
2458123.1 & 48.4 & 14.02 & 0.03 & 14.06 & 0.03 & 13.00 & 0.03 & 12.81 & 0.02 & 12.91 & 0.03 & 13.02 & 0.02\\
\hline
\end{tabular}
\label{table:UVOT}
\end{sidewaystable}

\begin{table*}
\centering
\caption{Optical spectroscopic and spectropolarimetric observations log.}
\begin{tabular}{rrrrrrr}
\hline
 Telescope & Instrument & date & $ t - t_0$ & Exposure time & Resolution & $\lambda$ Range   \\ 
&  & & (days) &  (s) & & ($\mathrm{\AA}$)\\
\hline
SOAR & Goodman & 2017-11-27 & 10  & 450 & 1000 & 3800\,--\,7800\\
Magellan & MIKE-Red & 2017-12-03 & 16 & TBA & TBA & 4800\,--\,9400\\
 Ir$\acute{\mathrm{e}}$n$\acute{\mathrm{e}}$e Du Pont & Echelle spec. & 2017-12-10 & 23 & 1200 & TBA & 3600\,--\,9000\\
 Ir$\acute{\mathrm{e}}$n$\acute{\mathrm{e}}$e Du Pont & Echelle spec. & 2017-12-11 & 24 & 1200 & TBA & 3600\,--\,9000\\
Magellan & MagE & 2017-12-15 & 28 & 600 & 4100 & 3500\,--\,9500\\
SMARTS & CHIRON & 2017-12-19 & 32 & 1800 & 25000 & 4100\,--\,8900\\
SALT & SpecPol & 2017-12-22 & 35 & 1200 & 5000 & 4000\,--\,7200\\
SMARTS & CHIRON & 2017-12-28 & 41 & 1800 & 25000 & 4100\,--\,8900\\
Magellan & MagE & 2017-12-30 & 43 & 600 & 4100 & 3500\,--\,9500\\
 Ir$\acute{\mathrm{e}}$n$\acute{\mathrm{e}}$e Du Pont & WFCCD & 2018-01-13 & 57 & 600 & 1000 & 3800\,--\,9500\\
SOAR & Goodman & 2018-02-11 & 86 & 300 & 1000 & 3800\,--\,7800\\
SOAR & Goodman & 2018-02-11 & 86 & 1200 & 5000 & 4500\,--\,5150\\
SOAR & Goodman & 2018-02-21 & 96 & 300 & 1000 & 3800\,--\,7800\\
SOAR & Goodman & 2018-02-21 & 96 & 1200 & 5000 & 4500\,--\,5150\\
SOAR & Goodman & 2018-03-14 & 117 & 600 & 1000 & 3800\,--\,7100\\
SOAR & Goodman & 2018-03-14 & 117 & 600 & 3000 & 4200\,--\,5500\\
SOAR & Goodman & 2018-03-24 & 127 & 900 & 5000 & 4500\,--\,5150\\
SOAR & Goodman & 2018-04-14 & 148 & 600 & 1000 & 3800\,--\,7800\\
SOAR & Goodman & 2018-04-14 & 148 & 600 & 5000 & 4500\,--\,5150\\
SOAR & Goodman & 2018-05-04 & 168 & 600 & 1000 & 3800\,--\,7800\\
SOAR & Goodman & 2018-05-04 & 168 & 900 & 5000 & 4500\,--\,5150\\
SOAR & Goodman & 2018-05-07 & 171 & 1200 & 1000 & 3800\,--\,7800\\
SOAR & Goodman & 2018-06-08 & 203 & 1200 & 1000 & 3800\,--\,7800\\
SOAR & Goodman & 2018-06-18 & 213 & 1200 & 1000 & 3800\,--\,7800\\
SOAR & Goodman & 2018-06-18 & 213 & 1200 & 5000 & 4500\,--\,5150\\
SOAR & Goodman & 2018-07-09 & 234 & 1200 & 1000 & 3800\,--\,7800\\
SOAR & Goodman & 2018-07-22 & 247 & 1200 & 1000 & 3800\,--\,7800\\
SOAR & Goodman & 2018-07-28 & 253 & 1200 & 1000 & 3800\,--\,7800\\
SOAR & Goodman & 2018-08-14 & 270 & 900 & 1000 & 3800\,--\,7800\\
SOAR & Goodman & 2018-08-14 & 270 & 1620 & 5000 & 4500\,--\,5150\\
SALT & SpecPol & 2018-10-05 & 322 & 1200 & 5000 & 4000\,--\,7200\\
\hline
\end{tabular}
\label{table:spec_log}
\end{table*}



\end{document}